\documentclass[11pt,a4paper]{article}
\pdfoutput = 1

\usepackage{jcappub}
\usepackage[utf8]{inputenc}
\usepackage{amsmath,amssymb}
\usepackage{graphicx}
\usepackage{hyperref,orcidlink}
\usepackage{natbib}
\usepackage{xcolor}
\usepackage{physics}
\usepackage{derivative}
\usepackage{subcaption}
\usepackage[T1]{fontenc}
\usepackage{orcidlink}

\newcommand{\SC}[1]{\textcolor{magenta}{{\bf SC:~{#1}}}}
\newcommand{\ZV}[1]{\textcolor{red}{{ZV:~{#1}}}}

\newcommand{\edit}[1]{{#1}}

\newcommand{\bq}{\textbf{q}}
\newcommand{\bp}{\textbf{p}}
\newcommand{\bx}{\textbf{x}}
\newcommand{\bk}{\textbf{k}}

\newcommand{\br}{\textbf{r}}

\newcommand{\half}{\frac{1}{2}}
\newcommand{\third}{\frac{1}{3}}

\def \dtq{\int d^3 \bq \ }

\newcommand{\avg}[1]{\ensuremath{\left\langle #1 \right\rangle}}

\def\Mpccube{\, h^{-3} \, {\rm Mpc}^3}

\def\kMpc{\, h \, {\rm Mpc}^{-1}}

\def\hq{\hat{q}} 

\def\hk{\hat{k}} 

\def\hp{\hat{p}}
\def\PL{P_{\rm L}}

\newcommand*{\lb}  {\left(}
\newcommand*{\rb}  {\right)}
\newcommand*{\la}  {\left\langle}
\newcommand*{\ra}  {\right\rangle}

\newcommand{\vq} {\vec{q}}

\newcommand{\vk} {\vec{k}}

\newcommand{\eq}[1]{\begin{align}#1\end{align}}
\newcommand{\eeq}[1]{\begin{equation}#1\end{equation}}
\newcommand*{\df}  {\delta}
\newcommand*{\non}  {\nonumber}
\renewcommand{\vec}{\textbf}

\subheader{
\footnotesize
\vspace*{-3.7em}
\begin{flushright}
RBI-ThPhys-2024-09
\end{flushright}
}

\title{The Bispectrum in Lagrangian Perturbation Theory}

\author[a]{Shi-Fan Chen\orcidlink{0000-0002-5762-6405}}
\author[b,c,d]{Zvonimir Vlah\orcidlink{0000-0002-9274-5768}}
\author[e]{Martin White\orcidlink{0000-0001-9912-5070}}

\affiliation[a]{School of Natural Sciences, Institute for Advanced Study, 1 Einstein Drive, Princeton, NJ 08540}
\emailAdd{sfschen@ias.edu}
\affiliation[b]{Division of Theoretical Physics, 
Ru\dj er Bo\v{s}kovi\'{c} Institute, Zagreb HR-10000, Croatia,}
\affiliation[c]{Kavli Institute for Cosmology Cambridge, Madingley Road, Cambridge, CB3 0HA, UK,}
\affiliation[d]{DAMTP, Centre for Mathematical Sciences, Wilberforce Road, Cambridge CB3 0WA, UK.}
\emailAdd{zvlah@irb.hr}
\affiliation[e]{Department of Physics, University of California,
Berkeley, CA, USA}
\emailAdd{mwhite@berkeley.edu}

\abstract{
We study the bispectrum in Lagrangian perturbation theory. Extending past results for the power spectrum, we describe a method to efficiently compute the bispectrum in LPT, focusing on the Zeldovich approximation, in which contributions due to linear displacements are captured to all orders in a manifestly infrared (IR) safe way. We then isolate the effects of these linear displacements on oscillatory components of the power spectrum like baryon acoustic oscillations or inflationary primordial features and show that the Eulerian perturbation theory (EPT) prescription wherein their effects are resummed by a Gaussian damping of the oscillations arise as a saddle-point approximation of our calculation. These two methods of IR resummation are in excellent agreement at 1-loop in the bispectrum. At tree level, resummed EPT does less well to capture the nonlinear damping of the oscillations, and the LPT calculation does not require an artificial split of the power spectrum into smooth and oscillatory components, making the latter particularly useful for modeling exotic features. We finish by extending our analysis of IR resummation in LPT to N-point functions of arbitrary order.
}
\begin{document}

\maketitle

\section{Introduction}
\label{sec:intro}

The observed large-scale structure is thought to arise through the action of gravitational instability in a cold dark matter dominated, expanding Universe from (very nearly) Gaussian initial conditions \cite{Dodelson20}.  The pattern of non-Gaussianity thus imparted is of a very particular form, and the tower of cumulants of the density and velocity fields on large scales can be systematically computed within perturbation theory \cite{Ber02}.  While for a Gaussian field the power spectrum is a sufficient statistic, and thus contains all of the primordial information, once the field becomes non-Gaussian there is information to be gained about gravitational interactions, cosmological parameters and bias relations from measurements beyond the power spectrum. For a given set of galaxy bias parameters cosmological perturbation theory makes consistent predictions for all of these n-point functions, which can be combined with the power spectrum in data analyses to either improve power-spectrum-only constraints or access new physical signals not present in the 2-point function. Recent years have in particular seen rapid advances in the use of the 3-point bispectrum to measure primordial non-Gaussianity \cite{Cabass22,DAmico23,Cabass24} and tighten constraints on cosmological parameters \cite{Ivanov23}. N-point functions are particularly appealing because, on scales where the non-Gaussianity is weak and well-controlled, the available cosmological signal is predominantly at low N such that the tower of cumulants can be truncated at some finite order.  In fact, recent work suggests that measurement of the low-order cumulants contains effectively all of the perturbative information that is present in the galaxy density field if attention is restricted to the larger scales where the modeling is most robust, unless the contributions of certain modes is accidentally parametrically large compared to expectations at a given order in perturbation theory \cite{Cabass24}.

The presence of a feature in the linear theory power spectrum due to baryon acoustic oscillations \cite{Sunyaev70,Peebles70} is one such complication in the perturbative calculation of N-point functions, since it leads to a ``small'' parameter related to the large-scale displacements of galaxies that is numerically quite large.  In order to handle this it is necessary to sum a subset of the full perturbative terms to high order, a procedure that goes by the name of ``IR resummation.''  Since it fundamentally traces the advection of matter due to long-wavelength perturbations, IR resummation is very naturally handled within the Lagrangian formulation of perturbation theory (LPT; \cite{Zeldovich70,Buchert87,Bouchet95,Matsubara08a,Matsubara08b,CLPT,VlaSelBal15,Chen21}), and indeed most implementations of IR resummation\edit{, including those grounded in the Eulerian framework of fluid dynamics (EPT), use Lagrangian arguments} at some point in their formulation (see e.g.\ \cite{SenZal15,Baldauf15} and the discussion in \cite{VlaWhiAvi15}).  

Beyond the direct Lagrangian resummation of long-wavelength displacements it is also possible to specifically isolate the BAO feature using a ``wiggle no-wiggle split'', estimate the effect of long-displacements on it, and resum only those particular effects. Since this procedure resums the linear displacements only in the wiggle component we will call it \texttt{RWiggle} for short throughout this paper. The result is effectively a Gaussian damping of BAO wiggles in the linear power spectrum \cite{ESW07} and can be shown using a simple saddle-point approximation of the full Lagrangian calculation \cite{Vlah16} or by diagramatically selecting enhanced diagrams in time-sliced perturbation theory (TSPT) \cite{Blas16}. The agreement between these two procedures has been very well studied for the 2-point function, i.e.\ the power spectrum, and the level of agreement of different schemes is well understood \cite{ChenVlahWhite20a}. A similar study has not yet been performed for cumulants beyond second order, with essentially all calculations to date using the Gaussian damping ansatz due in part to its simplicity compared to the full Lagrangian calculation for statistics beyond the power spectrum.

Beyond the theoretical advantage of being able to compare different IR resummation schemes, the full Lagrangian calculation also has the advantage that it does not require that oscillatory features in the initial conditions be artificially separated. This separation is not inherent to the underlying physical theory and, while well-studied in the case of the BAO, will need to be done on a case-by-case basis if, for example, searching for oscillations in the power spectrum due to exotic inflationary models. The IR resummation of the wiggles can still be performed in this case \cite{Vasudevan19}, but depends on more conditions such as distinguishing between smooth and oscillatory components when there is no single BAO-like scale in the problem, making the out-of-the box nature of the Lagrangian calculation an appealing alternative. This will be particularly useful for upcoming spectroscopic surveys, which will enable us to place extremely tight bounds on these inflationary features \cite{Chen10,Chluba15,Planck18,Slosar19,Beutler19}, and for extending the treatment of IR resummation to oscillations with complicated shapes in the bispectrum to constrain e.g.\ interactions of the inflaton with massive particles with spin (also known as the ``cosmological collider'') \cite{Cabass18,Dizgah18,Cabass24} .

The purpose of this paper is to investigate how the bispectrum and higher order functions can be efficiently computed within LPT while fully resumming the effect of long-wavelength displacements, and what we learn about IR resummation from considering the LPT bispectrum.  We are not the first to compute n-point functions in LPT for this purpose: ref.~\cite{Tassev14b} for example considered the configuration-space 3-point function in the Zeldovich approximation by brute force, multi-dimensional integration while ref.~\cite{Rampf12} formulated the LPT bispectrum in the spirit of integrated perturbation theory (iPT) \cite{Matsubara08a}. However, we update upon earlier work and extend the calculation of the 3-point function to Fourier space using fast \texttt{FFTLog}-based numerical techniques and study the phenomenology of oscillations in the bispectrum in detail, fully taking into account mode-coupling terms and elucidating the connection between the full Lagrangian calculation and the \texttt{RWiggle} prescription. The structure of the paper is as follows: After describing the general structure and expression for the bispectrum in LPT in \S\ref{sec:LPTbispectrum}, we show how it can be efficiently evaluated in an IR-safe way cancelling unwanted divergences in \S\ref{sec:direct_integration}. In \S\ref{sec:ir_resummation} we describe the behavior of BAO wiggles in some detail, comparing \texttt{RWiggle} with the direct Lagrangian calculation via numerical techniques derived in the previous section, as well as making contact with other IR resummation schemes in the literature. This discussion is extended to n-point functions of arbitrary order in \S\ref{sec:npoint}. We conclude in \S\ref{sec:conclusions} and discuss the straightforward extension to nonlinear bias and redshift-space distortions as well as an example calculation of the damping of inflationary features with a logarithmic dependence on wavenumber. Various technical details are derived in the Appendices. Numerical calculations throughout this paper will be done assuming the Planck 2018 best-fit flat $\Lambda$CDM cosmology \cite{Planck18} ($\Omega_m = 0.3111$, $h = 0.6766$, $\omega_b =  0.02242$, $n_s =  0.9665$, $\sigma_8 = 0.8102$) at redshift $z=0$ unless otherwise indicated.

\section{The Bispectrum in Lagrangian Perturbation Theory}
\label{sec:LPTbispectrum}

Lagrangian perturbation theory (LPT; \cite{Zeldovich70,Buchert87,Bouchet95,Matsubara08a,Matsubara08b,CLPT,VlaSelBal15}) is by now well established and efficient codes for numerical computation of power spectra exist \cite{ChenVlahWhite20a}.  We will not repeat all of the details here, instead referring the reader to the above-cited articles; briefly LPT models the growth of structure by following the displacements of fluid elements from their initial, Lagrangian positions ($\mathbf{q}$) to their final, Eulerian positions ($\mathbf{x}$) via a displacement: $\mathbf{x}=\mathbf{q}+\mathbf{\Psi}(\mathbf{q})$.  The displacement, $\mathbf{\Psi}$, is then systematically expanded order by order in the initial density perturbation.  Biased tracers are modeled by introducing a functional, $F$, of the initial density, velocity and tidal shear and their low-order derivatives. Within this formalism the galaxy density $\delta_g(\bx)$ has a Fourier transform given by \cite{Matsubara08a,Matsubara08b}
\begin{equation}
    \delta_g(\bk) = \dtq e^{-i\bk \cdot \bq} \Big( F(\bq) e^{-i\bk \cdot \Psi(\bq)} - 1 \Big)\, .
    \label{eqn:deltak}
\end{equation}
where $F(\mathbf{q})$ indicates the bias functional evaluated at the Lagrangian coordinate $\mathbf{q}$. For convenience we will ignore the $-1$ piece below, since it is nonzero only for vanishing wavenumbers and is anyway trivial to evaluate. As cosmological perturbation theory is an \textit{effective} theory modeling the large-scale fluid limit of structure formation, both the bias functional $F(\bq)$ and the perturbative expansion of the displacement $\Psi(\bq)$ have contributions whose sizes are free parameters, allowing us to marginalize over unknown small scale effects like galaxy formation (see e.g. \cite{PorSenZal14,VlaWhiAvi15,Des16}); since we focus on the role of long-wavelength linear displacements in this work we will not further discuss this aspect of LPT, and PT in genereal, except where relevant, leaving interested readers to consult the literature on this well-studied topic (see e.g. refs.~\cite{Baldauf14,Baldauf21} for thorough treatments of the matter bispectrum at 1- and 2-loops).

From the above the bispectrum can then be written as
\begin{equation*}
    \avg{\delta(\bk_1) \delta(\bk_2)  \delta(\bk_3)} = \int_{\bq_1,\bq_2,\bq_3}  e^{-i\bk_1 \cdot \bq_1-i\bk_2 \cdot \bq_2-i\bk_3 \cdot \bq_3} \avg{F_1 F_2 F_3\ e^{-i\bk_1 \cdot \Psi_1-i\bk_2 \cdot \Psi_2-i\bk_3 \cdot \Psi_3}} \, ,
\end{equation*}
where we have used the shorthand $f_n = f(\bq_n).$\footnote{For clarity of presentation, we will adopt the shorthands
\begin{equation}
    \int_{\bq_n} \equiv \int d^3\bq_n\, , \quad \int_{\bp_n} \equiv \int \frac{d^3\bp_n}{(2\pi)^3}\, ,
\end{equation}
for configuration and Fourier space integrals, respectively. We also also label Lagrangian quantities by their coordinates, e.g.\ $F_n = F(\bq_n)$.
}  This expression can be simplified by noting that the expectation value \edit{has to be} translation invariant \edit{by statistical homogeneity}, i.e.\ does not depend on all three Lagrangian coordinates $\bq_{1,2,3}$ independently. Adopting the coordinate system $(\bq, \br, \textbf{Q}) = (\bq_1-\bq_3,\bq_2-\bq_3,\bq_3)$, we can therefore eliminate the $\textbf{Q}$ dependence in the bracketed mean and write
\begin{align*}
    \avg{\delta(\bk_1) \delta(\bk_2)  \delta(\bk_3)} &= \int_{\bq,\br,\textbf{Q}} e^{-i\bk_1 \cdot \bq -i\bk_2 \cdot \br -i\bk_{123} \cdot \textbf{Q} }  \avg{F(\bq) F(\br) F(\textbf{0})\ e^{-i\bk_1 \cdot \Psi(\bq)-i\bk_2 \cdot \Psi(\br)-i\bk_3 \cdot \Psi(\textbf{0})}} \\
    &= (2\pi)^3 \delta_D(\bk_{123})\ \int_{\bq,\br} e^{-i\bk_1 \cdot \bq -i\bk_2 \cdot \br } \avg{F_1 F_2 F_3\ e^{-i\bk_1 \cdot \Psi_1-i\bk_2 \cdot \Psi_2 + i (\bk_1 + \bk_2) \cdot \Psi_3}} \, ,
\end{align*}
where $\bk_{123} = \bk_1 + \bk_2 + \bk_3$, and in the last line, we have used that this correlator is zero unless $\bk_{123} = 0$. The bispectrum is defined to be this correlator without the $\delta$-function, i.e. \cite{Rampf12}
\begin{equation}
    B(\bk_1,\bk_2) = \int_{\bq,\br} e^{-i\bk_1\cdot\bq-i\bk_2\cdot\br} \avg{F_1 F_2 F_3\ e^{-i\bk_1 \cdot \Delta_{13} -i\bk_2 \cdot \Delta_{23}  } }_{(\bq,\br)=(\bq_1-\bq_3\bq_2-\bq_3)} \, ,
    \label{eqn:bispectrum}
\end{equation}
where the pairwise displacement is $\Delta_{nm} = \Psi_n - \Psi_m$. Importantly, the right-hand side of Equation~\ref{eqn:bispectrum} is invariant under uniform shifts of the displacement, i.e. satisfies Galilean symmetry.

\subsection{Matter Bispectrum in the Zeldovich Approximation}
\label{sec:BZel}

Let us start in the case of matter in the Zeldovich approximation \cite{Zeldovich70}, i.e.\ lowest order LPT.  Then $F_n = 1$ and $\Psi = \Psi^{(1)}$ with $\Psi^{(1)}(\bk)=i(\bk/k^2)\delta_{\rm lin}$.  In the Zeldovich approximation displacements are Gaussian, since they are first order in $\delta_{\rm lin}$, so the cumulant theorem allows us to write
\begin{equation}
  \avg{e^{-i\bk_1 \cdot \Delta_{13} -i\bk_2 \cdot \Delta_{23}  } }
  = \exp\Big\{-\half \Big( k_{1,i} k_{1,j} \avg{\Delta_{13}^i \Delta_{13}^j}+k_{2,i} k_{2,j}
  \avg{\Delta_{23}^i \Delta_{23}^j} + 2 k_{1,i} k_{2,j} \avg{\Delta_{13}^i \Delta_{23}^j} \Big)\Big\}\, . \nonumber
\end{equation}
The moments of the pairwise displacements can be written as functions of the pairwise separations $\bq$, $\br$ by translation invariance \cite{CLPT} 
\begin{equation*}
    \avg{\Delta_{13}^i \Delta_{13}^j} = A_{ij}(\bq)\, , \quad \avg{\Delta_{23}^i \Delta_{23}^j} = A_{ij}(\br) \, ,
\end{equation*}
and can be decomposed into tensor comonents as $A_{ij}(\bq) = X(q) \delta_{ij} + Y(q) \hq_i \hq_j$ \cite{Matsubara08a}. The ``mixed'', third term is slightly more complicated; it can be written as 
\begin{align}
    \avg{\Delta_{13}^i \Delta_{23}^j} &= \avg{(\Psi_1 - \Psi_3)^i (\Psi_2 - \Psi_3)^j} \nonumber \\
    &= \avg{\Psi_1^i \Psi_2^j} - \avg{\Psi_1^i \Psi_3^j} - \avg{\Psi_3^i \Psi_2^j} + \avg{\Psi_3^i \Psi_3^j} \nonumber \\
    &= \left(\avg{\Psi_1^i \Psi_2^j} - \avg{\Psi^i \Psi^j}\right) + \left(\avg{\Psi^i \Psi^j} - \avg{\Psi_1^i \Psi_3^j}\right) + \left(\avg{\Psi^i \Psi^j} - \avg{\Psi_3^i \Psi_2^j} \right) \nonumber \\
    &= \half \left[  A_{ij}(\br) + A_{ij}(\bq) -A_{ij}(\bq_{12}) \right];\quad \bq_{12} = \bq_1 - \bq_2 = \bq - \br\,
    \label{eqn:DeltaiNDeltajN}
\end{align}
\edit{where $\avg{\Psi^i \Psi^j}$ is square displacement evaluated at a signle point}. The full expectation value can thus be written as
\begin{align}
& \avg{e^{-i\bk_1 \cdot \Delta_{13} -i\bk_2 \cdot \Delta_{23}  } } \nonumber \\
&= \exp\Big\{-\half \Big( k_{1,i} k_{1,j} A_{ij}(\bq)+k_{2,i} k_{2,j} A_{ij}(\br) + k_{1,i} k_{2,j} \big(   A_{ij}(\bq) + A_{ij}(\br) - A_{ij}(\bq-\br)\big) \Big)\Big\} \nonumber \\
&= \exp\Big\{\half \Big( k_{1,i} k_{3,j} A_{ij}(\bq)+k_{2,i} k_{3,j} A_{ij}(\br) + k_{1,i} k_{2,j}  A_{ij}(\bq-\br)\Big)\Big\}\, ,
\label{eqn:zel_exponent}
\end{align}
where in the last line, we have used momentum conservation $\bk_1 + \bk_2 + \bk_3 = 0$ to restore the symmetry between the sides of the triangle. Note that this form is symmetric in both the wavenumbers $\bk_{1,2,3}$ and the three vertices of the triangle $\bq_{1,2,3}$.

\subsection{Tree Level}
\label{ssec:tree}
It will be quite instructive to check equation~\ref{eqn:zel_exponent} at the lowest order, i.e.\ tree level, in such a way as to make contact with the more familiar Eulerian perturbation theory (EPT) calculation. In the Zeldovich approximation, expanding order-by-order in the initial conditions, the density can be written as
\begin{equation*}
    \delta(\bk) = \sum_{n=0}^\infty \int_{\bp_1, ..., \bp_n} \ Z_n(\bp_1, ... , \bp_n)\ \delta_0(\bp_1) ... \delta_0(\bp_n)\ (2\pi)^3 \delta_D(\bk-\bp_{\rm tot}) \, ,
\end{equation*}
where \cite{Grinstein87}
\begin{equation}
    Z_n(\bp_1, ... , \bp_n) =  \frac{1}{n!} \left( \frac{\bp_\textrm{tot} \cdot \bp_1}{p_1^2} \right) \cdots \left( \frac{\bp_\textrm{tot} \cdot \bp_n}{p_n^2} \right)\, .
\end{equation}
In particular, we have that
\begin{equation}
    Z_2(\bp_1,\bp_2) = \frac{1}{2} \Big( 1 + \frac{\bp_1 \cdot \bp_2}{p_1^2} \Big) \Big( 1 + \frac{\bp_1 \cdot \bp_2}{p_2^2} \Big)\, ,
\end{equation}
such that the tree-level bispectrum of the Zeldovich matter field is given by
\begin{equation}
    B^{\rm tree}_{\rm Zel}(\bk_1,\bk_2,\bk_3) = 2\, Z_2(\bk_1,\bk_2)\, \PL(k_1) \PL(k_2) + 2\, \textrm{cycle} \, .
\label{eqn:BtreeZelEPT}
\end{equation}
Let us compare this to Equation \ref{eqn:zel_exponent}. To get a nonzero Fourier transform, we have to start at the second order term when Taylor expanding the exponential
\begin{align}
    B^{\rm tree}_{\rm Zel} =& \int_{\bq,\br} e^{-i\bk_1\cdot\bq-i\bk_2\cdot\br}\ \Big[ \Big( \half k_{1,i} k_{3,j} A_{ij}(\bq) \Big)\Big( \half k_{2,i} k_{3,j} A_{ij}(\br) \Big)  \\
    &+ \Big( \half k_{1,i} k_{3,j} A_{ij}(\bq) \Big)\Big( \half k_{1,i} k_{2,j} A_{ij}(\bq-\br) \Big) + \Big( \half k_{2,i} k_{3,j} A_{ij}(\br) \Big)\Big( \half k_{1,i} k_{2,j} A_{ij}(\bq-\br) \Big) \Big]\, . \nonumber
\end{align}
The quadratic terms shown here are symmetric under permutations of the triangle coordinates in coordinate and Fourier space, so we need only look at the first term, which is the simplest due to being factorizable in the coordinates we have chosen with $\bq_3$ `at the origin'. In fact, this first piece gives
\begin{equation}
    \int_{\bq} e^{-i\bk_1\cdot\bq}\ \Big( \half k_{1,i} k_{3,j} A_{ij}(\bq) \Big) \int_{\br} e^{-i\bk_2\cdot\br}\ \Big( \half k_{2,i} k_{3,j} A_{ij}(\br) \Big)
    = 2\, Z_2(\bk_1,\bk_2)\, \PL(k_1) \PL(k_2) \, ,
\end{equation}
where in obtaining this result we have used that $\bk_3 = -(\bk_1+\bk_2)$.  Transforming the latter two pieces of $B^{\rm tree}_{\rm Zel}$ give the cyclically permuted results so the full result is in agreement with equation \ref{eqn:BtreeZelEPT}. Note that evaluating the Fourier transform above technically also incurs terms that are $\textit{linear}$ in the power spectrum, e.g.
\begin{equation*}
    \int_{\bq,\br} e^{-i\bk_1 \cdot \bq-i\bk_2 \cdot \br}\ \left( \frac12 k_{1,i,} k_{3,j} A_{ij}(\bq) \right) = \left( \frac{ \bk_1 \cdot \bk_3 }{k_1^2} \right) \PL(k_1) (2\pi)^3 \delta_D(\bk_2) \, ,
\end{equation*}
as well as a zeroth order term equal to two $\delta$ functions; this is because we have ignored the $-1$ in Equation~\ref{eqn:bispectrum} which cancels out these disconnected pieces.

\section{Direct IR-Safe Resummation of Zeldovich Displacements}
\label{sec:direct_integration}

While LPT and EPT agree order by order, as we showed in the previous section at tree level, the nonlinear structure of the exponentiated displacements LPT has physical consequences beyond this level, particularly for capturing the nonlinear damping of the BAO, which we will return to in \S\ref{sec:ir_resummation}. It is thus of great interest to develop efficient numerical techniques to evaluate the double integral in Equation~\ref{eqn:bispectrum}.

\subsection{Preliminary Steps: Generalizing the Zeldovich Power Spectrum}
\label{ssec:epsilon}

Motivated by previously developed techniques to efficiently compute the power spectrum within LPT, we introduce
\eeq{
\mathcal E_{12}(\vec p) \equiv \mathcal E(\vec k_1, \vec k_2, \vec p) = \int_{\vec q} e^{-i \vec p \cdot \vec q}\, e^{+\frac{1}{2} \vk_{1,i} \vk_{2,j}  A_{ij}(\vec q)}\, , 
~~{\rm i.e.}~~
e^{+\frac{1}{2} \vk_{1,i} \vk_{2,j}  A_{ij}(\vec q)} = \int_{\vec p}\, e^{i \vec p \cdot \vec q}\, \mathcal E_{12}(\vec p)\, ,
\label{eqn:epsilon}
}
which is equal to the Zeldovich power spectrum when $\bp = \bk_1 = -\bk_2$ (i.e.\ for $\bk_1 + \bk_2 = 0$).  Note the $+$ sign in the exponent containing $A_{ij}$. The objects $\mathcal{E}_{nm}(\bp)$ are the Fourier transforms of the three exponentiated factors that make up the cumulant in Equation~\ref{eqn:zel_exponent}. Techniques for evaluating this Fourier transform in that special case were developed in ref.~\cite{VlaSelBal15}, and we extend them to allow for arbitrary $\bk_{1,2}$ and $\bp$ in Appedix~\ref{app:epsilon_numerics}; roughly, while in the case of the power spectrum all three vectors $\bk_{1,2}$ and $\bp$ were co-linear, here they can form arbitrary spatial configurations, such that the result is a function of both the individual lengths $k_1, k_2, p$ as well as the angles between the vectors specified by their dot products $\hk_1 \cdot \hk_2, \hk_1 \cdot \hp, \hk_2 \cdot \hp$. We find that accounting for these new degrees of freedom, for a given bispectrum configuration specified by $\bk_{1,2}$, the integral can be performed for all $\bp$ in terms of the product of two matrices, one consisting of radial Hankel transforms, which can be efficiently computed using the \texttt{FFTLog} algorithm \cite{Ham00}, and another depending solely on angular polynomial coefficients.

An important point is that since the integrand in Equation~\ref{eqn:epsilon} does not asymptote to zero as $\bq\to \infty$, $\mathcal{E}$ will have both a finite part and one proportional to a $\delta$-function, i.e.
\begin{equation}
    \mathcal{E}_{12}(\bp) = \mathcal{E}_{12}^{\rm fin}(\bp) + (2\pi)^3 \delta_D(\bp)\ e^{ \half (\bk_1 \cdot \bk_2) \Sigma^2} \, ,
    \label{eqn:epsilon_split_sigma}
\end{equation}
where $3\Sigma^2$ is the mean pairwise displacement of two points separated by $\bq \rightarrow \infty$. While we focus on computing the numerically tricky finite piece in Appendix~\ref{app:epsilon_numerics}, keeping track of both pieces is important for the stability of the integral, as we will discuss below. A particularly useful way to re-express Equation~\ref{eqn:epsilon_split_sigma} is
\begin{align}
    \mathcal{E}_{12}(\bp) = \mathcal{E}^{\rm fin}_{12}(\bp) + \left(1 - \int_{\bp'} \mathcal{E}^{\rm fin}_{12}(\bp') \right) (2\pi)^3 \delta_D(\bp)\, .
    \label{eqn:epsilon_split_integral}
\end{align}
This can be seen by noting that, since $A_{ij}(\bq \rightarrow 0) = 0$, the $\bq=0$ Fourier transform of $\mathcal{E}$ must be equal to unity and is again a consequence of Galilean invariance.

\subsection{An IR-safe Integrand for the Bispectrum}
\label{ssec:ir_safe_integrand}

Since the Zeldovich bispectrum is the Fourier transform of three exponentiated $A_{ij}$'s, it can be written using Equation \ref{eqn:zel_exponent} in terms of their Fourier transforms $\mathcal{E}$ as
\eq{
B(\bk_1,\bk_2) = 
\int_{\bq_1, \bq_2}\, e^{-i \vk_1 \cdot \vq_1 - i \vk_2 \cdot \vq_2} \la e^{- i \vk_1 \cdot \Delta_1 - i \vk_2 \cdot\Delta_2}\ra 
%= \int_{\vec p}\, \mathcal E(\vec k_1, \vec k_3, \bk_1 - \vec p) \mathcal E(\vec k_2, \vec k_3,  \bk_2 + \vec p)  \mathcal E(\vec k_1, \vec k_2, \vec p)\, .
= \int_{\vec p}\,
\mathcal E_{13}(\bk_1 - \vec p) 
\mathcal E_{23}(\bk_2 + \vec p) 
\mathcal E_{12}(\vec p)\, .
\label{eqn:epsilon_to_b}
}
The configuration space integral above represents a double Fourier transform, which is difficult to evaluate numerically and thus has represented an obstacle in any practical implementation of bispectrum LPT results. Conversely, recasting it in the momentum space, the integral becomes a straightforward convolution amenable to direct integration, given the techniques to compute $\mathcal{E}$ in the previous subsection. Indeed, by splitting up the finite and $\delta$-function pieces, we obtain
\begin{align}
\int_{\vec p}\,
\mathcal E_{13}(\bk_1 - \vec p) 
\mathcal E_{23}(\bk_2 + \vec p) 
\mathcal E_{12}(\vec p) 
%& = \int_{\vec p}\, \mathcal E^{\rm fin}(\vec k_1, \vec k_3, \bk_1 - \vec p)  \mathcal E^{\rm fin}(\vec k_2, \vec k_3,  \bk_2 + \vec p) \mathcal E^{\rm fin}(\vec k_1, \vec k_2, \vec p) + e^{(\bk_1\cdot\bk_2) \Sigma^2} \mathcal E_{\rm fin}(\vec k_1, \vec k_3, \bk_1) \mathcal E_{\rm fin}(\vec k_2, \vec k_3,  \bk_2) \nonumber \\ &+ e^{(\bk_1\cdot\bk_3) \Sigma^2} \mathcal E_{\rm fin}(\vec k_2, \vec k_3,  \bk_3) \mathcal E_{\rm fin}(\vec k_1, \vec k_2, \bk_1) + e^{(\bk_2\cdot\bk_3) \Sigma^2} \mathcal E_{\rm fin}(\vec k_1, \vec k_2, \bk_2) \mathcal E_{\rm fin}(\bk_1, \bk_3, \bk_3)
= \,
&e^{\half (\bk_1\cdot\bk_2) \Sigma^2} \mathcal E^{\rm fin}_{13}(\bk_1) \mathcal E^{\rm fin}_{23}(\bk_2) + 2\,{\rm cycle} \non \\
& + \int_{\vec p}\, 
\mathcal E^{\rm fin}_{13}(\bk_1 - \vec p) 
\mathcal E^{\rm fin}_{23}(\bk_2 + \vec p) 
\mathcal E^{\rm fin}_{12}(\vec p) \, ,
\label{eqn:integral_ir_unsafe}
\end{align}
where we have dropped any disconnected pieces requiring one of the external momenta $\bk_i$ to be zero. However, as we shall show now, it is possible to re-absorb these $\delta$-funtion contributions into the integral to obtain a numerically better-behaved integral.

The above integral, if implemented numerically as written, is rather sensitive to the behavior of the integrand terms when the argument of $\mathcal{E}$'s approaches zero, where they have a pole. As a particularly simple example, let us first look at the linearized result, where
\begin{equation}
    \mathcal{E}^{\rm lin}_{12}(\bp) = -\frac{(\bk_1 \cdot \bp)(\bk_2 \cdot \bp)}{p^4} \PL(p) + \left(1 + \half (\bk_1 \cdot \bk_2) \Sigma^2 \right) (2\pi)^3 \delta_D(\bp) \, ,
\end{equation}
that has a pole at $p = 0$. In the long-wavelength (infrared) region where $p < k_{\rm IR} \ll k_{1,2}$ we get the contribution
\begin{equation}
    \left[ 1 + k_{1,i} k_{2,j} \left( \half \Sigma^2 \delta_{ij} - \int_{|\bp| < k_{\rm IR}} \frac{p_i p_j}{p^4} \PL(p)  \right) \right]\ \mathcal{E}_{\rm lin}(\bk_1, \bk_3, \bk_1)  \mathcal{E}_{\rm lin}(\bk_2, \bk_3, \bk_2)\, ,
    \label{eqn:ir_limit}
\end{equation}
where ``1'' piece is simply the tree level result derived in \S\ref{ssec:tree}. More interestingly, in the parentheses we see that the infrared contribution to Zeldovich square displacement $\Sigma^2$ from the $\delta$-function piece is exactly cancelled by the equivalent infrared contribution by convolving the finite pieces, leading to no net effect. However, as written in Equation~\ref{eqn:integral_ir_unsafe}, this cancellation depends on the agreement between a convolution for the finite piece and direct integration of the linear power spectrum for the $\delta$-function piece, which is not guaranteed. While the $\bp$ integral in Equation~\ref{eqn:ir_limit} is not formally divergent in $\Lambda$CDM, its contribution on typical scales for LSS analyses typically has $k^2 \Sigma^2 \gtrsim 1$, requiring us to rely on the cancellation of two numbers in Equation~\ref{eqn:integral_ir_unsafe}.

\begin{figure}
    \centering
    \includegraphics[width=1.0\textwidth]{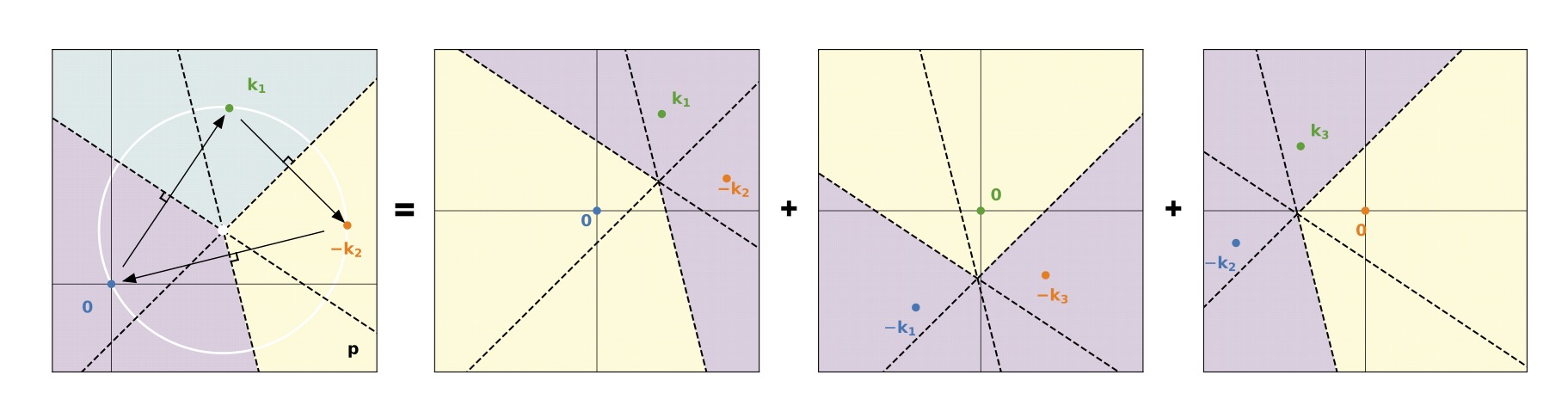}
    \caption{The bispectrum integral separated into three domains according to Equation~\ref{eqn:domains}. Each domain is defined to by proximity to one of the infrared poles of the integral ${\bf 0}, \bk_{1}, -\bk_{2}$, such that they are separated by the perpendicular bisectors of a triangle with sidelengths $k_{1,2,3}$ in the plane formed by the poles, and join at the circumcenter. By performing coordinate transformations separately in each domain such that the pole is sent to the origin, we can rewrite the bispectrum as a sum of three integrals, each with a domain defined by the $k_{1,2,3}$ triangle shifted to a different position such that a different vertex is at the origin. We see also that in addition to $\pm \bk_{1,2}$ some of these triangles also develop poles at $\pm \bk_{3}$, restoring the permutation symmetry of the bispectrum.}
    \label{fig:ir_regions}
\end{figure}

In order to build the infrared cancellation more explicitly into the bispectrum, it will be convenient to slightly rearrange the arguments in Equation~\ref{eqn:epsilon_to_b}. This is because the $\mathcal{E}$'s in the integrand have arguments $\bp, \bk_1 - \bp, \bk_2 + \bp$, such that the integrand has ``infrared'' regions about three points ${\bf 0}, \bk_1$ and $-\bk_2$. However, much like in the case of the power spectrum \cite{Carrasco14} and 1-loop bispectrum \cite{Baldauf14} it is possible to isolate each singularity in the integrand and map them onto the origin to yield a more symmetric expression. Specifically, we make use of the fact that the locus of points closest to each pole partitions the space into three non-overlapping regions to split the integral as
\begin{align}
    \int_\bp  &= \int_{p\ <\ |\bk_1 - \bp|,\ |\bk_2 + \bp|}+ \int_{|\bk_1 - \bp|\ <\  p,\ |\bk_2 + \bp|} + \int_{|\bk_2 + \bp|\ <\ |\bk_1 - \bp|,\  p} \nonumber \\
    &= \int_{p\ <\ |\bk_1 - \bp|,\ |\bk_2 + \bp|}+ \int_{p'\ <\  |\bk_1 - \bp'|,\ |\bk_3 + \bp'|} + \int_{p''\ <\ |\bp'' -\bk_2|,\ |\bk_3 + \bp''|}\, .
    \label{eqn:domains}
\end{align}
In the second line we have for the latter two integrals performed the coordinate transformations $\bp' = \bk_1 - \bp$ and $\bp'' = \bk_2 + \bp$. \edit{As shown in Figure~\ref{fig:ir_regions}, this} corresponds to a Voronoi tessellation of the poles in the integrand and can be extended also to higher N-point functions (\S\ref{sec:npoint}); the above decomposition was also used in EPT by ref.~\cite{Baldauf14} for the $B_{222}$ contribution to the 1-loop bispectrum.\footnote{See also ref.~\cite{Angulo15} who use a slightly different domain decomposition to remove the same IR poles.} We see that each of the singularities in the bispectrum integrand exists in one of these partitions, which we have mapped onto $\bp, \bp', \bp'' = \bf{0}$, such that in each domain the closest singularity is the one at the origin. These domains have boundaries set by the perpendicular bisectors formed by the triangle formed by the poles with side lenghts $k_{1,2,3}$ and meet at the circumcenter. Rewriting the above in terms of \edit{Heaviside} $\Theta$ functions, the bispectrum integral is now
\begin{align}
    B(\bk_1, \bk_2) = \int_\bp &\mathcal{E}_{12}(\bp)\ \mathcal{E}_{13}(\bk_1 - \bp)\ \mathcal{E}_{23}(\bk_2 + \bp)\ \Theta(|\bk_1 - \bp| - p)\ \Theta(|\bk_2 + \bp| - p)  \nonumber \\
    +\ \  &\mathcal{E}_{12}(\bk_1 - \bp)\ \mathcal{E}_{13}(\bp)\ \mathcal{E}_{23}(-\bk_3 - \bp)\ \Theta(|\bk_1 - \bp| - p)\ \Theta(|\bk_3 + \bp| - p)\nonumber \\ 
    +\ \  &\mathcal{E}_{12}(-\bk_2 + \bp)\ \mathcal{E}_{13}(-\bk_3 - \bp)\ \mathcal{E}_{23}(\bp)\ \Theta(|\bk_2 - \bp| - p)\ \Theta(|\bk_3 + \bp| - p) \, .
    \label{eqn:symm_integrand}
\end{align}
%where we have used the shorthand $\mathcal{E}_{nm}(\bp) = \mathcal{E}(\bk_n,\bk_m,\bp)$. 
This integral can be made manifestly symmetric in $\bk_{1,2,3}$ by using that $\mathcal{E}_{nm}$ is an even function of $\bp$, but we will not perform this additional step here for the sake of brevity.

We can now write this integrand in an IR safe way by explicitly re-introducing the $\delta$-function piece. Plugging Equation~\ref{eqn:epsilon_split_integral} back into the bispectrum integral (Eq.~\ref{eqn:epsilon_to_b}) we have
\begin{align}
     B(\bk_1, \bk_2) &= \int_{\bp} \mathcal{E}_{12}^{\rm fin}(\bp) \mathcal{E}_{13}(\bk_1 - \bp) \mathcal{E}_{23}(\bk_2 + \bp) + \left(1 - \int_\bp \mathcal{E}_{12}^{\rm fin}(\bp) \right) \mathcal{E}_{13}(\bk_1) \mathcal{E}_{23}(\bk_2) \nonumber \\
     &= \mathcal{E}_{13}(\bk_1) \mathcal{E}_{23}(\bk_2) + \int_{\bp} \mathcal{E}_{12}^{\rm fin}(\bp) \left( \mathcal{E}_{13}(\bk_1 - \bp) \mathcal{E}_{23}(\bk_2 + \bp) - \mathcal{E}_{13}(\bk_1) \mathcal{E}_{23}(\bk_2)\right)\, .
     \label{eq:Bis_IRsafe}
\end{align}
Both pieces of the integral are now IR safe: the product in front is since it is derived purely from pairwise displacements, while the integrand in the second piece vanishes as $p \rightarrow 0$. From this we can see that the full bispectrum integrand is given by
\begin{align}
    B(\bk_1, \bk_2) &=\ \mathcal{E}_{13}^{\rm fin}(\bk_1) \mathcal{E}_{23}^{\rm fin}(\bk_2) + \mathcal{E}_{13}^{\rm fin}(\bk_3) \mathcal{E}_{12}^{\rm fin}(\bk_2) + \mathcal{E}_{12}^{\rm fin}(\bk_1) \mathcal{E}_{23}^{\rm fin}(\bk_3) \nonumber \\
    &+ \int_\bp \Big\{ \mathcal{E}_{12}^{\rm fin}(\bp) \mathcal{E}_{13}^{\rm fin}(\bk_1 - \bp) \mathcal{E}_{23}^{\rm fin}(\bk_2 + \bp)\ \Theta(|\bk_1 - \bp| - p) \Theta(|\bk_2 + \bp| - p)  \nonumber \\
    &+ \ \ \mathcal{E}^{\rm fin}_{12}(\bk_1 - \bp) \mathcal{E}^{\rm fin}_{13}(\bp) \mathcal{E}_{23}^{\rm fin}(-\bk_3 - \bp)\ \Theta(|\bk_1 - \bp| - p) \Theta(|\bk_3 + \bp| - p)\nonumber \\ 
    &+ \ \ \mathcal{E}_{12}^{\rm fin}(-\bk_2 + \bp) \mathcal{E}_{13}^{\rm fin}(-\bk_3 - \bp) \mathcal{E}_{23}^{\rm fin}(\bp)\ \Theta(|\bk_2 - \bp| - p) \Theta(|\bk_3 + \bp| - p) \nonumber \\
    &-\ \ \mathcal{E}_{12}^{\rm fin}(\bp) \mathcal{E}_{13}^{\rm fin}(\bk_1) \mathcal{E}_{23}^{\rm fin}(\bk_2) - \mathcal{E}_{23}^{\rm fin}(\bp) \mathcal{E}_{13}^{\rm fin}(\bk_3) \mathcal{E}_{12}^{\rm fin}(\bk_2) - \mathcal{E}_{13}^{\rm fin}(\bp)\mathcal{E}_{12}^{\rm fin}(\bk_1) \mathcal{E}_{23}^{\rm fin}(\bk_3) \Big\} \, ,
    \label{eqn:ir_safe}
\end{align}
such that the poles in each region picked out by the $\Theta$ functions are explicitly cancelled by the poles at the origin in the last line.

\begin{figure}
    \centering
    \includegraphics[width=0.95\textwidth]{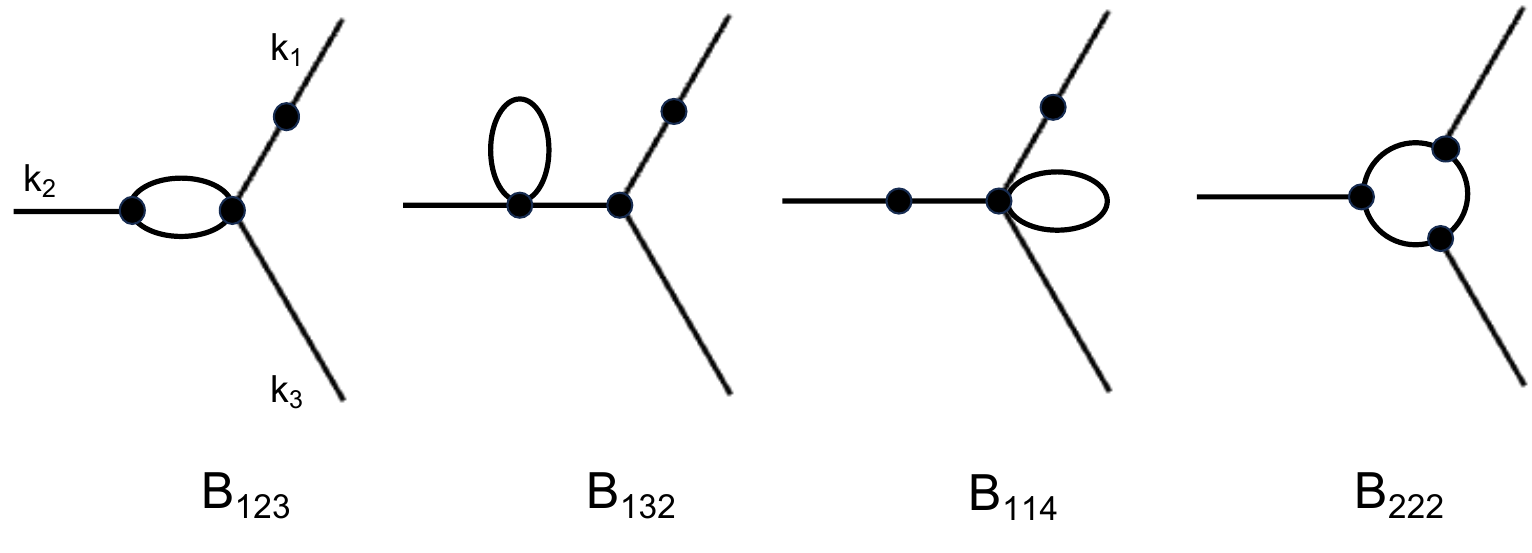}
    \caption{Contributions to the 1-loop bispectrum. Each of the first three diagrams is required to cancel the IR contribution to the fourth.}
    \label{fig:loop_diagrams}
\end{figure}

The above also enables us to write down an infrared-safe expression for the 1-loop Zeldovich bispectrum. Splitting $\mathcal{E}^{\rm fin} = \mathcal{E}^{(1)} + \mathcal{E}^{(2)} + ...$ into its $\mathcal{O}(\PL^n)$ contributions we have that the tree level bispectrum is obtained by the replacement to the first line of Equation~\ref{eqn:ir_safe} with $B_{\rm tree} \sim \mathcal{E}^{(1)} \mathcal{E}^{(1)}$ while the 1-loop contributions are given by the contributions $B_{\rm 1-loop} \sim \mathcal{E}^{(1)} \mathcal{E}^{(2)}$ and $\int_\bp \mathcal{E}^{(1)} \mathcal{E}^{(1)} \mathcal{E}^{(1)}$. Appendix~\ref{app:epsilon_numerics} details how the order $n$ contributions to $\mathcal{E}$ can be isolated in numerical calculations and efficiently calculated. Since the infrared safety of Equation~\ref{eqn:ir_safe} holds order by order, the equivalent expressions here at 1-loop are manifestly infrared safe as well. This is in contrast to the IR-safe integrands for the 1-loop EPT bispectrum developed in ref.~\cite{Baldauf14,Angulo15} where the integrals of all the loop diagrams have to be combined after different remappings to cancel IR divergences. In Appendix~\ref{app:ept_1loop} we diagrammatically review the 1-loop contributions to the bispectrum, showing why unlike in our method where each piece is individually IR safe each of the diagrams shown in Figure~\ref{fig:loop_diagrams} contain IR contributions which have to be combined together to cancel in an IR-safe way.

\subsection{Numerical Implementation}

We are now in a position to numerically investigate the Zeldovich bispectrum. In order to do so we have developed \texttt{triceratops}\footnote{\url{https://github.com/sfschen/triceratops}} a \texttt{Python} code that computes the Zeldovich bispectrum in various approximations (tree-level, 1-loop, full resummation) using the methods described in this section. \texttt{triceratops} computes $\mathcal{E}$ efficiently using the \texttt{FFTLog} algorithm \cite{Ham00} and performs the convolution integral using a three-dimensional, spherical grid in $\bp$ that is logarithmically spaced in $p$ and use Gaussian weights for $\theta$ and $\phi$. As an additional check, we also developed in parallel a \texttt{Mathematica} code to perform the same calculations using both the in-built integration routines as well as the \texttt{vegas} algorithm \cite{vegas}---while not as fast as \texttt{triceratops}, the two codes are in excellent agreement, showing that the underlying integrals are well-behaved.

\begin{figure}
    \centering
    \includegraphics[width=\textwidth]{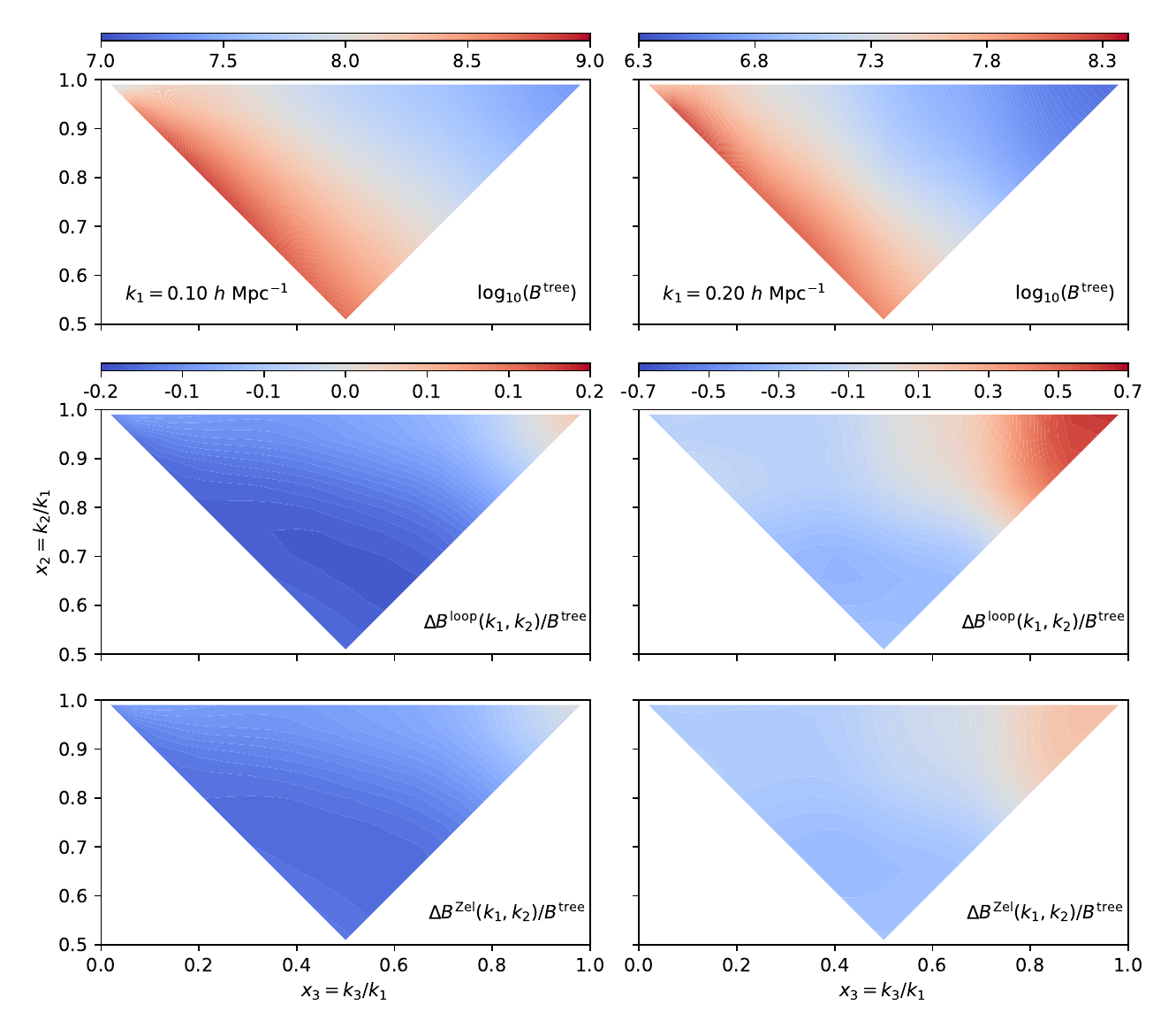}
    \caption{The tree level, 1-loop Zeldovich and full Zeldovich matter bispectra for triangles with longest sidelengths of $k = 0.10$ (left) and $k = 0.20 \kMpc$ (right) at $z=0$. Each row shows the logarithm of the tree level bispectrum in $\Mpccube$ units (top) and the ratios of the \edit{1-loop (middle) and full Zeldovich (bottom) corrections ($\Delta B = B - B^{\rm tree}$) to the tree-level bispectrum.}}
    \label{fig:triangle_plots}
\end{figure}

\begin{figure}
    \centering
    \includegraphics[width=\textwidth]{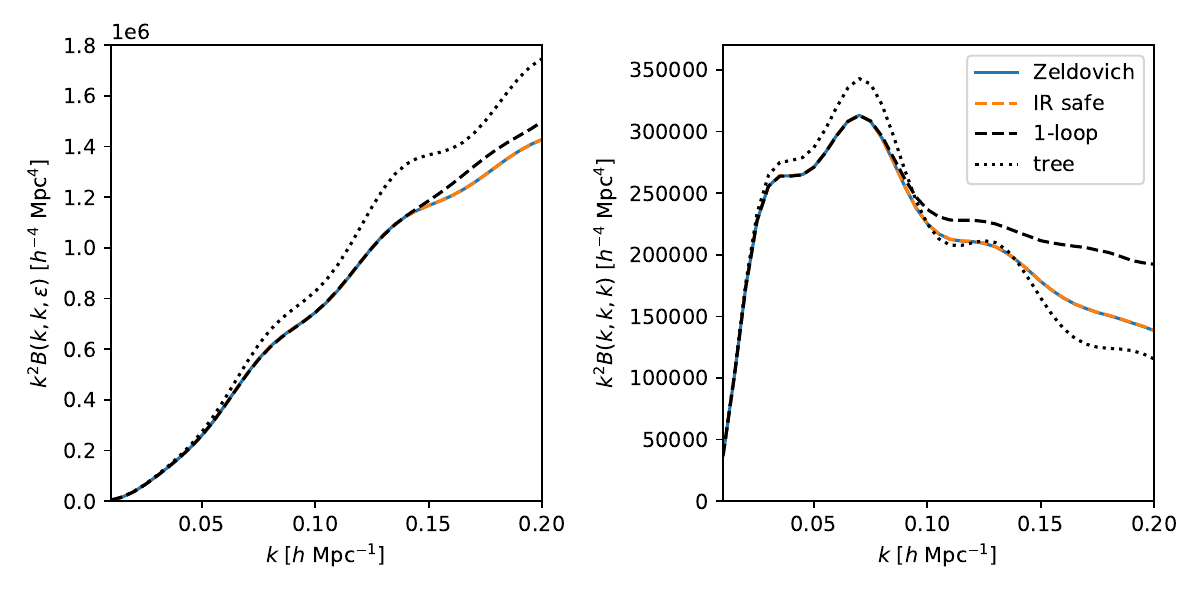}
    \caption{The $z=0$ matter Zeldovich bispectrum in the squeezed (left) and equilateral configurations (right), computed using the methods described in \S\ref{sec:direct_integration}, as a function of side length. The \edit{solid blue} and \edit{orange dashed} lines show the calculation performed using Equation~\ref{eqn:integral_ir_unsafe} and the IR-safe Equation~\ref{eqn:ir_safe}, respectively, which are in excellent agreement when the numerics are properly converged. The dashed line shows the 1-loop bispectrum, which agrees with the full Zeldovich calculation at low $k$. The squeezed configuration is taken to be one where the short side is $5\%$ that of the longer sides.}
    \label{fig:numerical}
\end{figure}

Figures~\ref{fig:triangle_plots} and \ref{fig:numerical} show the bispectrum computed using the methods described above in the squeezed and equilateral configurations. Reassuringly, both the full and 1-loop Zeldovich calculations approach the leading-order tree-level calculation at low wavenumbers, while differing at higher ones, with the 1-loop and full Zeldovich calculations agreeing with each other to slightly higher wavenumbers than the tree-level ones. 

\begin{figure}
    \centering
    \includegraphics[width=\textwidth]{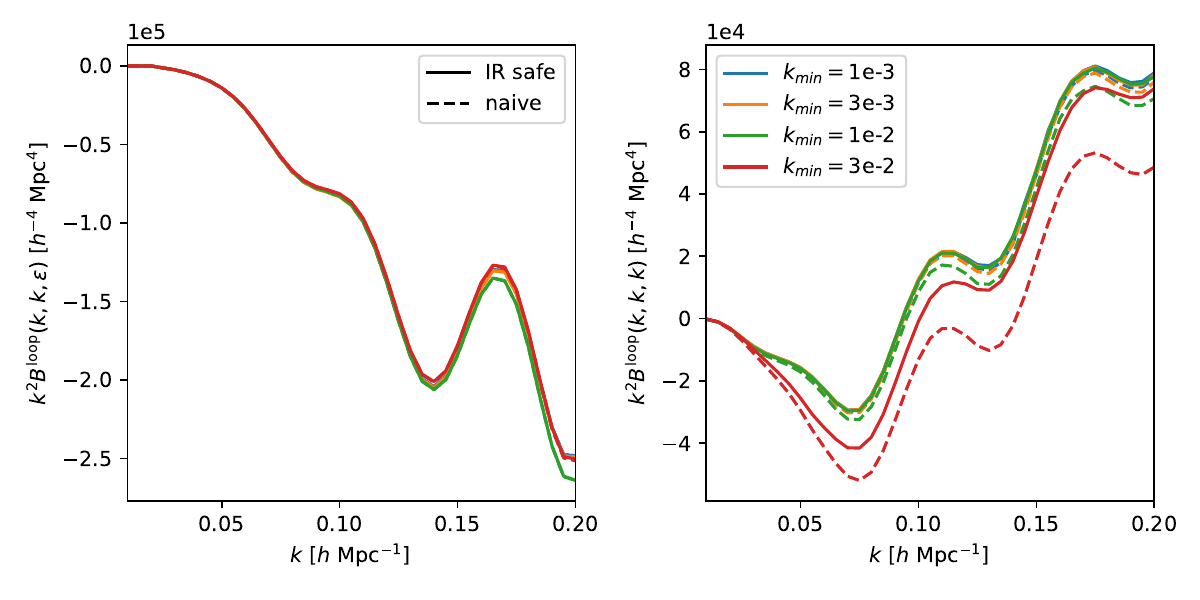}
    \caption{Numerical stability of the bispectrum calculation to the minimum wavenumber sampled in the integral $k_{\rm min}$, in the squeezed (left) and equilateral (right) configurations. Dashed and solid lines show the 1-loop contribution to the Zeldovich bispectrum using the naive and IR-safe integrands, respectively, with the latter converging to the correct result faster at higher $k_{\rm min}$ as expected. }
    \label{fig:kmin1loop}
\end{figure}

Figure~\ref{fig:numerical} also shows the full Zeldovich calculation using both the naive (Eqn.~\ref{eqn:integral_ir_unsafe}) and IR safe (Eqn.~\ref{eqn:ir_safe}) integrands. Both are in excellent agreement for the settings shown in the plot. However, in order to compare their behavior when the sampling of small wavenumbers is varied, we compute the 1-loop contribution to the bispectrum in Figure~\ref{fig:kmin1loop} using both methods---here it is quite clear that the IR-safe integrand is far less sensitive to the scales sampled, recovering the correct result even when the smallest wavenumber sampled $k_{\rm min}$ is quite close to the wavenumber arguments in the bispectrum itself. This is of course as expected since the IR-safe integrand has explicitly vanishing contributions from long-wavelength modes. Finally, we have also tested the sensitivity of our bispectrum calculation to the maximum wavenumber $k_{\rm max}$ sampled in the integral, as well as the number of Hankel transforms required when calculation $\mathcal{E}$---we find that the algorithm is relatively lenient in these requirements, since the dependence on $\mathcal{E}$ is not very spread out in Fourier space but rather localized to physically-relevant wavenumbers. In particular, each $\mathcal{E}$ is sufficiently captured by including up to five spherical-bessel transforms, much like is typically needed for the power spectrum on perturbative scales, as long as the bispectrum is evaluated also on these scales.

\section{Infrared Resummation of BAO and the Saddle-Point Approximation}
\label{sec:ir_resummation}

\subsection{Review: Resummation for the Power Spectrum}

In the preceding sections we developed the formalism to compute the bispectrum of galaxies in Lagrangian perturbation theory with the linear displacements fully resummed. In practice, it is often desirable to look specifically at the effect of these displacements on the BAO feature, where their effects are most enhanced. In order to do so it is useful to isolate the BAO feature in the linear spectrum by performing a ``wiggle no-wiggle split'' $\PL = \PL^{nw} + f_b \PL^{w}$ \cite{ESW07} where $f_b \PL^w$ is a purely oscillatory component carrying the BAO while $\PL^{nw}$ is the smooth component of the power spectrum. We have included the baryon fraction $f_b$ as a counting parameter since we will be interested primarily in working at first order in the BAO, but we will keep it only implicitly below, with the understanding that each superscript $w$ carries a factor of $f_b$. The definition of $\PL^{nw}$ is inexact and many algorithms to compute it exist in the literature with differing results (see e.g. \cite{Chen24} for a review of commonly used methods). Throughout the numerical portions of this work we will adopt the B-spline method in ref.~\cite{Vlah16}, which has the nice property that the total variance in linear displacements is preserved, though we note that the exact choice is somewhat a matter of taste and an inherent systematic of calculations involving this split.

The most prominent effect of the exponentiated linear displacements in the power spectrum is to produce a roughly Gaussian damping of the BAO feature, which can therefore be resummed by including this Gaussian factor whenever the wiggle piece of the power spectrum appears (\texttt{RWiggle}). This effect has long been known empirically in the large-scale structure literature \cite{ESW07,Crocce08,Matsubara08a} and was derived analytically in perturbation theory as an infrared resummation of long-wavelength modes enhanced by the BAO by refs. \cite{SenZal15,Baldauf15,Vlah16,Blas16}. An equivalent derivation, presented in ref.~\cite{Vlah16}, shows that this form can be simply derived by performing a saddle-point approximation about the BAO in Lagrangian perturbation theory, and we briefly review it below.

Using the wiggle no-wiggle split we can separate $A_{ij} = A^{nw}_{ij} + A^w_{ij}$ into two pieces due to the wiggle and no-wiggle power spectra
\begin{equation}
    A_{ij}^{w,nw}(\bq) = 2 \int_\bk \left( 1 - e^{i\bk\cdot\bq} \right) \left( \frac{k_i k_j}{k^4} \right) \PL^{w,nw}(k)\, .
\end{equation}
The former will be strongly peaked at the BAO scale $q = r_s$ while the latter will be essentially a smooth version of $A_{ij}$ without the BAO feature. Performing this split we have that the Zeldovich matter power spectrum for matter, at first order in the wiggles, is
\begin{align}
    P_{\rm Zel}^w(k) &= -\frac12 k_i k_j \int_{\vec q} e^{i\bk\cdot\bq - \frac12 k_i k_j A^{nw}_{ij}(\bq)} A^w_{ij}(\bq) \nonumber\\
    &\approx -\frac12 k_i k_j e^{-\half k^2 X_s} \int_{\vec q} e^{i\bk\cdot\bq - \frac12 k_i k_j Y_s \hq_i \hq_j} A^w_{ij}(\bq)\nonumber \\
    &\approx e^{-\half k^2 \Sigma^2_{s}} P^w_{\rm L}(k),\quad k r_s \gg 1\, .
    \label{eqn:saddle_point_pk}
\end{align}
In going from the first to second lines, we have used the saddle-point\footnote{Here by saddle point we refer to the approximation where when integrating the product of two functions $f, g$, if $f$ is narrowly 
 peaked at $x_0$ we can have
 \begin{equation}
     \int dx\ f(x) g(x) = g(x_0) \int dx\ f(x) + \mathcal{O}(\sigma)
 \end{equation}
 where $\sigma$ is the width of $f$ (Appendix~\ref{app:saddle_point}).
 } approximation to substitute the values of $X^{nw}(q),\ Y^{nw}(q)$ at the BAO scale $r_s$, which we have denoted using subscripts $X_s$, $Y_s$, rather than integrate over their full $q$ dependence. Corrections to this approximation can be computed by expanding the smooth component about $r_s$, i.e. by Taylor expanding $e^{-\half k_i k_j A^{nw}_{ij}}$ in $\Delta q = q - r_s$, with the correction scaling as powers of the BAO width $r_D$ multiplying the mean square density contrast on BAO scales. We provide further details in Appendix~\ref{app:saddle_point}.

\begin{figure}
    \centering
    \includegraphics[width=\textwidth]{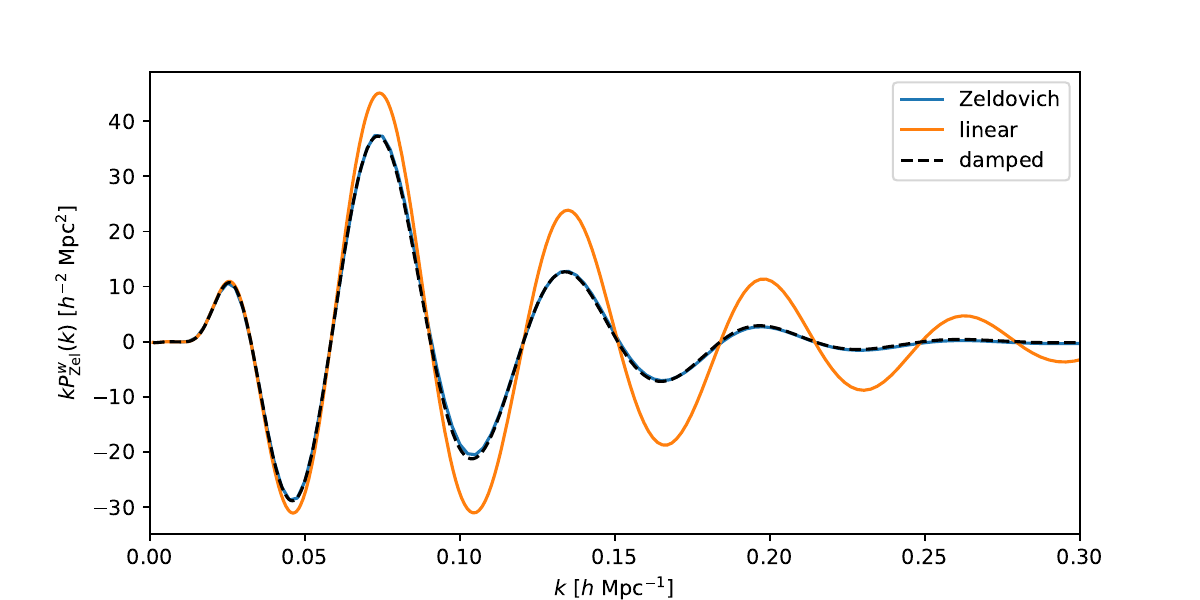}
    \caption{The wiggle component of the Zeldovich power spectrum, computed in full (blue), at linear order (orange) and in the \texttt{RWiggle} scheme. The Zeldovich prediction is suppressed relative to the linear one in a way that is very well described by a Gaussian damping.}
    \label{fig:pk_rwiggle}
\end{figure}

The remaining subtlety is the angular integral in the second line of Equation~\ref{eqn:saddle_point_pk}. In Appendix~\ref{sec:derivation_xpy}, we show that the angular integral in the penultimate line can be performed exactly in the limit that $k r_s \gg 1$ to yield the final expression with
\begin{equation}
    \Sigma^2_{s} = X_s + Y_s = \int \frac{dk}{3\pi^2}\ \PL(k)\ \left(1 - j_0(k r_s) + 2 j_2(k r_s) \right)\,
\end{equation}
\edit{where the $j_\ell$ are spherical Bessel functions}. At intermediate $k$ this approximation breaks down, leading to corrections to the simple exponential damping form. The $1 - j_0 + 2 j_2$ in the parentheses cuts off contributions from modes with wavelengths longer than the BAO scale, such that coherent displacements that move the BAO, i.e. act locally as a Galilean transformation, cannot damp it. As shown in Figure~\ref{fig:pk_rwiggle}, \texttt{RWiggle} describes the wiggles in the power spectrum almost perfectly even when everything else is kept to linear order. It is important to note that the above is a description of the parametrically large effects of linear displacements on the BAO feature---beyond them nonlinear effects and effective-theory corrections, which do not have to be limited to the Gaussian form above, can also play a role. These effects have been extensively studied in the perturbation-theory literature so we will not describe them in any detail in what follows, except to emphasize that even a complete model of the linear displacements does not fully specify the nonlinear shape of the wiggles.

We can also connect the saddle-point calculation above to the resummation developed in \S\ref{sec:direct_integration}. Specifically, we can rewrite the integral in the first line of Equation~\ref{eqn:saddle_point_pk} as
\begin{align}
    P_{\rm Zel}^w(k) %&= \dtq e^{i\bk\cdot\bq} \left\{ -\half k_i k_j A^w_{ij}(\bq) \times \exp\left( -\half k_i k_j A^{nw}_{ij}(\bq) \right) \right\} \nonumber \\
    &= \int_\bp \left\{ -\half k_i k_j \tilde{A}^w_{ij}(\bk-\bp) \times \text{FT}\left[\exp\left( -\half k_i k_j A^{nw}_{ij}(\bq) \right)\right](\bp) \right\} \nonumber \\
    &= \int_\bp \left\{ -\half k_i k_j \tilde{A}^w_{ij}(\bk-\bp) \times \mathcal E^{nw}(\bk,-\bk,\bp) \right\} %\label{eqn:conv} 
    \non \\
    &= e^{-\half k^2 \Sigma^2} P^w_{\rm L}(k) + \int_\bp \left\{ -\half k_i k_j \tilde{A}^w_{ij}(\bk-\bp) \times \mathcal E^{nw}_{\rm fin}(\bk,-\bk,\bp) \right\} \, ,
\label{eqn:wiggle_in_mometum_space}    
\end{align}
where $\mathcal E^{nw}$ is computed using the no-wiggle spectrum only and where in the last line we have used the finite-infinite split in Equation~\ref{eqn:epsilon_split_sigma} as required when numerically evaluating $\mathcal{E}$. The first term looks superficially similar to the damping form derived in Equation~\ref{eqn:saddle_point_pk}, with $\Sigma^2$ replacing $\Sigma^2_{s}$.
However, it would be incorrect and unphysical to assume that $P_{\rm Zel}^w$ is damped by the exponential in the $\delta$-function piece, which is what we would get if we dropped the finite piece. 

It is instructive to recover the result for the dampened wiggle obtained in Equation \ref{eqn:saddle_point_pk} starting from the momentum space representation given in Equation \ref{eqn:wiggle_in_mometum_space}. We first note that there are two potentially IR divergent contributions arising from the poles $\vec p\to 0$ and $\vec p \to \vec k$. The latter does not, in fact, contribute given that the wiggle power spectrum has no support in that regime. Thus, only the $\vec p\to 0$ pole remains and we can simply regularise the integral
\eq{
P_{\rm Zel}^w(k)
&= e^{-\half k^2 \Sigma^2} P^w_{\rm L}(k) 
+ \half k_i k_j \tilde{A}^w_{ij}(\bk) 
\int_\bp \mathcal E^{nw}_{\rm fin}(\bk,-\bk,\bp) \non\\
&\hspace{2.6cm} -\half k_i k_j 
\int_\bp \lb \tilde{A}^w_{ij}(\bk-\bp) - \tilde{A}^w_{ij}(\bk) \rb
\mathcal E^{nw}_{\rm fin}(\bk,-\bk,\bp) \non\\
&= P^w_{\rm L}(k) -\half k_i k_j 
\int_\bp \lb \tilde{A}^w_{ij}(\bk-\bp) - \tilde{A}^w_{ij}(\bk) \rb
\mathcal E^{nw}_{\rm fin}(\bk,-\bk,\bp)\, ,
}
where we used that $\int_\bp \mathcal E^{nw}_{\rm fin}(\bk,-\bk,\bp) = 1 - \exp \lb -k^2 \Sigma^2\rb$. This allows us to write an IR safe expression where we have explicitly cancelled the large bulk displacements, which do not therefore affect the shape of the wiggle power spectrum, as is expected from the equivalence principle and Galilean invariance. 
To derive the physical effect of long displacements on wiggles, we use the fact that the BAO bump is localized in the configuration space and write 
\begin{align}
    -\half k_i k_j  &\int_\bp \lb \tilde{A}^w_{ij}(\bk-\bp) - \tilde{A}^w_{ij}(\bk) \rb \ \mathcal E^{nw}_{\rm fin}(\bk,-\bk,\bp) \nonumber \\
    &= -\half k_i k_j  \int_{\bq,\bp} e^{i \bk \cdot \bq} \lb e^{i \bp \cdot \bq} - 1\rb A^w_{ij}(\bq) \mathcal E^{nw}_{\rm fin}(\bk,-\bk,\bp) \approx \lb e^{-\frac{1}{2}k^2 \Sigma^2_{s}} - 1\rb P^w_{\rm L}(k)\, .
\end{align}
%\eq{
%-\half k_i k_j  \int_\bp \lb \tilde{A}^w_{ij}(\bk-\bp) - \tilde{A}^w_{ij}(\bk) \rb
%&\mathcal E^{nw}_{\rm fin}(\bk,-\bk,\bp) \non\\
%&= -\half k_i k_j  \int_{\bq,\bp} e^{i \bk \cdot \bq} \lb e^{i \bp \cdot \bq} - 1\rb A^w_{ij}(\bq)
%\mathcal E^{nw}_{\rm fin}(\bk,-\bk,\bp) \non\\
%&\approx \lb e^{-\frac{1}{2}k^2 \Sigma^2_{s}} - 1\rb P^w_{\rm L}(k)\, .
%}
In order to resum only IR modes, we could have also explicitly restricted the $\vec p$ integral, i.e. modes for which we have $p\ll\Lambda\lesssim k$, which would have further restricted the contributions to $\Sigma_s$. As can be seen in the configuration and Fourier-space derivations above, this is not required at this stage to derive $\texttt{RWiggle}$ for the power spectrum. However, these IR modes come into play if we want to show that the resummation is under control and that the subleading contributions can indeed be neglected (see Appendixes~\ref{app:saddle_point} and \ref{sec:derivation_xpy}).

\subsection{Saddle-Point IR Resummation of the Bispectrum BAO }
\label{ssec:saddle}

\begin{figure}[t!]
    \centering
    \includegraphics[height=0.3\textheight]{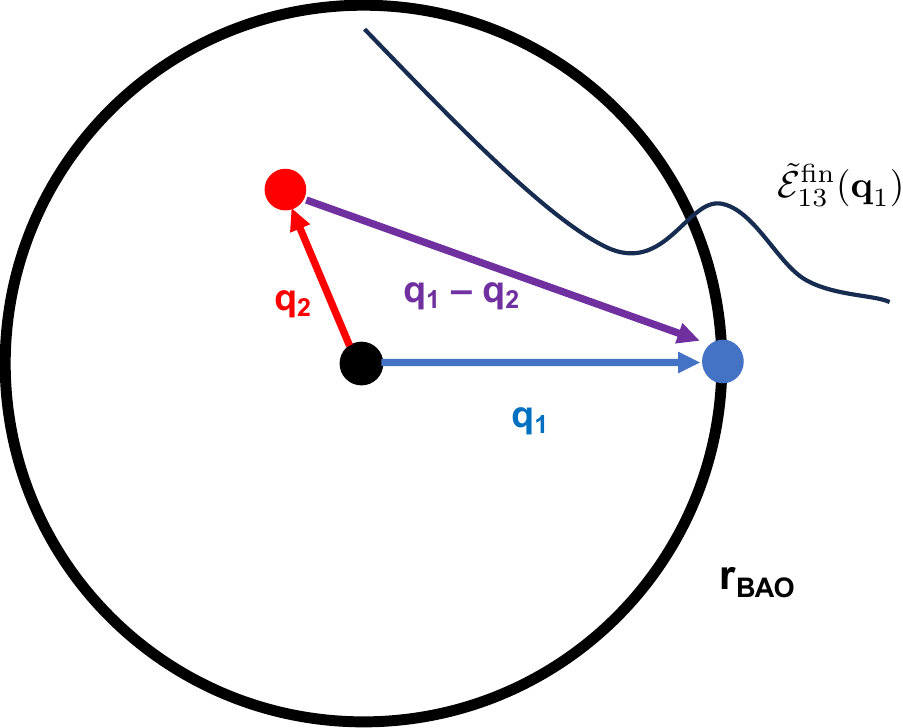}
    \caption{Triangle configuration relevant for the IR resummation of the BAO feature in Lagrangian space\edit{: two points (black and blue), or galaxies, are constrained to be separated by $\bq_1$ with $q_1 = r_{d}$ equal to the BAO radius, due to the localized BAO peak in $\tilde{\mathcal{E}}_{13}^{\rm fin}(\bq_1)$, while the third point (red dot) is unconstrained. The leading effect of long-wavelength modes on the BAO due to this additional degree of freedom can be obtained by expanding this third point about the origin.} }
    \label{fig:triangle_bao}
\end{figure}

%We can now extract the IR-resummed wiggle contributions to the bispectrum. The end result will be that the biggest effects of the long displacements can be captured by substituting in the \textit{damped} linear power spectrum (Eqn.~\ref{eqn:saddle_point_pk}) for the linear power spectrum. This was derived diagramatically by ref.~\cite{Blas16} using time-sliced perturbation theory (TSPT). Here we show that this result can be derived, as in the case of the 2-point function, using a saddle-point approximation, though additional corrections due to mode coupling appear that can be perturbatively managed. We will do so in two (related) ways, first by focusing on the contributions of long-wavelength modes in configuration space, then showing how the full linear Zeldovich displacements can be resummed in this way and the residuals computed in an IR-safe way.
Let us now extend the above analysis and extract the IR-resummed wiggle contributions to the bispectrum (\texttt{RWiggle}). The final result shows that the largest effects, due to long displacements, can be captured by linearized wiggle contributions to the linear power spectrum. We can derive this result, as in the case of the 2-point function, using a saddle-point approximation. We do so in two related ways, first (more restrictively) by focusing on the contributions of strictly long-wavelength modes in configuration space and subsequently relaxing this condition and showing how the full linear Zeldovich displacements can be resummed and the residuals computed in an IR-safe way. Our result, up to the linear terms in $P_{w}$, corresponds to the case where the effects of the long displacements are captured by substituting the \textit{damped} linear power spectrum (Eqn.~\ref{eqn:saddle_point_pk}) for the linear power spectrum. The latter result was derived diagrammatically by ref.~\cite{Blas16} using time-sliced perturbation theory (TSPT).

Let us begin, however, by considering the leading contribution expanding in the wiggle power spectrum $P_w$ to $\mathcal{E}$. Since each $A_{ij}$ in Equation~\ref{eqn:zel_exponent}, or each $\mathcal{E}$ in Equation~\ref{eqn:epsilon_to_b}, can have a power of $A_{ij}^w$ expanded out, we have three distinct contributions in total proportional to $P_{w}(k_{1,2,3})$. In particular, we can use the same logic as in the previous section to write each as
\begin{equation}
    \mathcal E^w(\bk_1,\bk_2,\bp) = \int_{\vec q} e^{i\bp\cdot\bq + \half k_{1,i} k_{2,j} A^{nw}_{ij}(\bq)} \left( \half k_{1,i} k_{2,j} A^w_{ij}(\bq) \right) \approx e^{\half (\bk_1 \cdot \bk_2) \Sigma^2_{s}} A(\bk_1, \bk_2, \bp) \PL^w(p)\, ,
\end{equation}
where we have defined the angular factor $A(\bk_n, \bk_m, \bp) = - (\bk_n \cdot \bp)(\bk_m \cdot \bp)/p^4$. Note that the damping of the BAO in $\mathcal{E}$ is itself IR safe, i.e. does not depend on displacements from modes longer than the BAO radius, since it is defined from pairwise displacements.

We can now proceed to consider BAO oscillations in the bispectrum itself. We will do this in two ways---first, by considering explicitly the effect of long-wavelength modes in (Lagrangian) configuration space and deriving the bispectrum version of \texttt{RWiggle} through a saddle-point calculation therein, and then by using an IR-safe integrand in Fourier space to extract the effect of all linear displacements and derive a convenient expression for evaluating the residual term.

\subsubsection{Infrared Modes and Configuration-Space Saddle Point}
\label{sssec:config_saddle}

We first consider the effect of purely infrared modes---that is, ones shorter than the wavenumbers entering a given bispectrum configuration, starting from Equation~\ref{eqn:ir_safe}. Specifically, in each of the three pieces of the integral we can extract contributions due to $\bp$ with $p$ smaller than $k_{1,2}/2$, $k_{1,3}/2$ or $k_{2,3}/2$. These regions are naturally set by the angular bisectors defining the IR regions. Focusing on one of the three domains without lack of generality, we have in the first one
\begin{align}
    (\text{IR\ Region of } k_{1,2}) &= \int_{|\bp|< k_{1,2}/2} \mathcal{E}_{12}^{\rm fin}(\bp) \left(  \mathcal{E}_{13}^{\rm fin}(\bk_1-\bp) \mathcal{E}_{23}^{\rm fin}(\bk_2+\bp) - \mathcal{E}_{13}^{\rm fin}(\bk_1) \mathcal{E}_{23}^{\rm fin}(\bk_2)\right) \nonumber \\
    &= \int_{\bq_{1,2}} \ e^{-i \bk_1 \cdot \bq_1 -i \bk_2 \cdot \bq_2}\ \left(\tilde{\mathcal{E}}^{\rm fin}_{12}(\bq_1 - \bq_2) - \tilde{\mathcal{E}}^{\rm fin}_{12}(\bf{0}) \right)_{\rm IR} \tilde{\mathcal{E}}^{\rm fin}_{13}(\bq_1) \tilde{\mathcal{E}}^{\rm fin}_{23}(\bq_2) \nonumber \\
    &= \int_{\bq_{1,2}} \ e^{-i \bk_1 \cdot \bq_1 -i \bk_2 \cdot \bq_2}\ \left( e^{\half k_{1,i} k_{2,j} A_{ij}(\bq_1-\bq_2)} - 1 \right)_{\rm IR} \tilde{\mathcal{E}}^{\rm fin}_{13}(\bq_1) \tilde{\mathcal{E}}^{\rm fin}_{23}(\bq_2)\, ,
\end{align}
where the subscript denotes that all the non-IR modes are removed, and in going from the second to third lines, we have used the integral relation in Equation~\ref{eqn:epsilon_split_integral}. We can now add in the ``tree-level'' piece, which cancels the ``minus one'':
\begin{align}
    \mathcal{E}^{\rm fin}_{13}(\bk_1) \mathcal{E}^{\rm fin}_{23}(\bk_2) &+ (\text{IR\ Region of } k_{1,2}) \nonumber \\
    &= \int_{\bq_{1,2}} \ e^{-i \bk_1 \cdot \bq_1 -i \bk_2 \cdot \bq_2}\ \left( e^{\half k_{1,i} k_{2,j} A_{ij}(\bq_1-\bq_2)} \right)_{\rm IR} \tilde{\mathcal{E}}^{\rm fin}_{13}(\bq_1) \tilde{\mathcal{E}}^{\rm fin}_{23}(\bq_2)\, . \label{eqn:config_space_saddle_integrand}
\end{align}
By construction, the BAO wiggle piece needs to come from the last two factors in the integrand and not the infrared first piece; splitting the two $\mathcal{E}$ into a wiggle and no-wiggle piece yields the tree level power spectrum contributions
\begin{align}
    \mathcal{E}_{13}^{w, \rm fin}(\bk_1) \mathcal{E}_{23}^{nw, \rm fin}(\bk_2) + & \mathcal{E}_{13}^{nw, \rm fin}(\bk_1) \mathcal{E}_{23}^{w, \rm fin}(\bk_2) \non\\
    &= Z_2(\bk_1, \bk_2) \left( P^w_{\rm L}(k_1) P^{nw}_{\rm L}(k_2) + P^{nw}_{\rm L}(k_1) P^w_{\rm L}(k_2) \right) + \mathcal{O}(\PL^3) \, ,
\end{align} 
yielding six total such contributions when all three IR domains of Equation~\ref{eqn:ir_safe} are taken into account.

We now consider the BAO saddle-point of Equation~\ref{eqn:config_space_saddle_integrand}. \edit{Expanding to first order in the wiggles, we see that the oscillatory contribution from $\mathcal{E}^w_{13}(\bq_1)$, which has support sharply peaked around $q_1 = r_d$, picks out the pieces of the integral looking like the configuration shown in Figure~\ref{fig:triangle_bao}. In particular, while one edge of the triangle corresponding to the Lagrangian separation $\bq_1$ has length fixed to the BAO scale, the length of the other two edges $\bq_2$ and $\bq_1 - \bq_2$ are free. This corresponds to the three-point function configuration where two of three galaxies are separated by the BAO scale. Unlike in the power spectrum case, which only involved separations of two galaxies at a fixed distance, we  have to contend with the exponential factor in Equation~\ref{eqn:config_space_saddle_integrand} having an extra degree of freedom in $\bq_2 - \bq_1$, which will in general induce additional dynamical corrections to the observed BAO along $\bk_1$ ($\bq_1$);} nonetheless, if we fix $\bq_2 = 0$ in the infrared piece, we get
\begin{align}
\mathcal{E}_{23}^{nw, \rm fin}(\bk_2) \int_{\bq_1} \ e^{-i \bk_1 \cdot \bq_1} \left( e^{\half k_{1,i} k_{2,j} A_{ij}(\bq_1)} \right)_{\rm IR} \tilde{\mathcal{E}}^{w, \rm fin}_{13}(\bq_1) 
&\approx e^{\half (\bk_1 \cdot \bk_2) \Sigma^2_{s}} \mathcal{E}_{13}^{w, \rm fin}(\bk_1) \mathcal{E}_{23}^{nw, \rm fin}(\bk_2)  \\
&\approx e^{-\half k_1^2 \Sigma^2_{s}} A(\bk_1, \bk_2, \bp) \PL^w(p) \mathcal{E}_{23}^{nw, \rm fin}(\bk_2)\, ,  \non
\end{align}
i.e. that the tree-level bispectrum is damped by IR displacements in a Gaussian way depending only on the wavenumber of the BAO wiggle itself.

The corrections due to this approximation come from the statistics of the other legs of the 3-point function triangle and can be computed as terms in a Taylor series
\begin{equation}
    \int_{\bq_2} e^{-i \bk_2 \cdot \bq_2} (\bq_{2,i_1} ... \bq_{2,i_n}) \ \tilde{\mathcal{E}}^{nw, \rm fin}_{23}(\bq_2)\int_{\bq_1} \ e^{-i \bk_1 \cdot \bq_1} \ \left(\frac{\partial^n}{\partial \bq_{1,i_1} ... \partial\bq_{1,i_n}} \right) \left( e^{\half k_{1,i} k_{2,j} A_{ij}(\bq_1)} \right)_{\rm IR} \tilde{\mathcal{E}}^{w, \rm fin}_{13}(\bq_1) \, , \nonumber
\end{equation}
where the first factor is simply the $\bk_2$ derivative of $\mathcal{E}^{nw,\rm fin}_{23}$. The second factor involves derivatives evaluated about $q_1 = r_d$; these take the form
\begin{equation}
    \left(\frac{\partial^n}{\partial \bq_{1,i_1} ... \partial \bq_{1,i_n}} \right) \left( e^{\half k_{1,i} k_{2,j} A_{ij}(\bq_1)} \right)_{q_1 = r_d} \sim \half k_{1,i} k_{2,j} \int_{\bp,\ p<1/r_d} \frac{p_i p_j p_{i_1} ... p_{i_n}}{p^4} \PL(p) \, .
    \label{eqn:ir_corrections}
\end{equation}
In a power-law universe where $\PL = A p^{n_s}$ it is easy to see that this integral scales as $r_d^{-n}$ with $n$, such that the n$^{\rm th}$ order contribution is suppressed by $(k_2 r_d)^{-n}$. We thus see that the damping can be captured by a simple damping of the input linear power spectrum, with perturbatively small corrections due to modes with wavelengths on the BAO scale.

\subsubsection{Fourier Space Resummation of Linear Displacements}
\label{sssec:fs_saddle}

Here, we show that the equivalent result as derived above can be obtained from the Fourier space representation of the bispectrum, allowing us to relax the requirement of restricting to the modes smaller than $k_{1,2}/2$.
We thus start with the expression for the bispectrum (still working in Zeldovich approximation) given in Equation \ref{eqn:epsilon_to_b}.
Since we are free to shift the integration variable, we can recover any relevant momenta combination by considering only the case when the wiggle contribution of $\mathcal E^{\rm w}$ appears as
\eq{
\int_{\vec p}\,
\mathcal E^{\rm nw}_{13}(\vec k_1 - \vec p) 
\mathcal E^{\rm nw}_{23}(\vec k_2 + \vec p) 
\mathcal E^{\rm w}_{12}(\vec p) \, .
}
Given that the remaining two contributions can be recovered by transforming $\vec p \to - \vec k_2 + \vec p$, and $\vec p \to \vec k_1 - \vec p$, we can write the linearized wiggle contribution to the bispectrum  as
\eq{
B^{w}
&= \int_{\vec p}\,
\mathcal E^{nw}_{13} (\vec k_1 - \vec p) 
\mathcal E^{nw}_{23} (\vec k_2 + \vec p) 
\mathcal E^{w}_{12} (\vec p)  + (2~{\rm cycle}) \non \\
&=e^{\half (\vec k_3 \cdot \vec k_1) \Sigma^2}
\Big( 
\mathcal E^{nw,{\rm fin}}_{32}(\vec k_3)  \mathcal E^{ }_{12}(\vec k_1)
+  \mathcal E^{w}_{32}(\vec k_3) \mathcal E^{nw,{\rm fin}}_{12}(\vec k_1) \Big) \non\\
&\hspace{4.5cm} +
\int_{\vec p}\,
\mathcal E^{nw,{\rm fin}}_{12} (\vec k_1 - \vec p) 
\mathcal E^{nw,{\rm fin}}_{23} (\vec k_2 + \vec p) 
\mathcal E^{w} (\vec k_1, \vec k_2, \vec p)
 + (2~{\rm cycle}) \non \\
&= 
\mathcal E^{nw,{\rm fin}}_{32}(\vec k_3)  \mathcal E^{w}_{12} (\vec k_1)
+ \mathcal E^{nw, {\rm fin}}_{23}(\vec k_2 ) \mathcal E^{\rm w}_{13} (\vec k_1) \non\\
&\hspace{4.5cm} +
\int_{\vec p, {\rm IR}}\,
\mathcal E^{nw, {\rm fin}}_{13} (\vec k_1 - \vec p) 
\mathcal E^{nw, {\rm fin}}_{23} (\vec k_2 + \vec p) 
\mathcal E^{w}_{12} (\vec p) 
 + (2~{\rm cycle})\, , 
\label{eq:B_wiggle_IR_safe_I}
}
where we have again used that $\int_{\vec p} \mathcal E^{\rm fin}_{12} (\vec p) =  1 - \exp(\half (\vec k_1 \cdot \vec k_2) \, \Sigma^2)$. In the above we also introduced the IR-regularized mode coupling integral
\eq{
&\int_{\vec p, {\rm IR}}\,
\mathcal E^{nw, {\rm fin}}_{13} (\vec k_1 - \vec p) 
\mathcal E^{nw,{\rm fin}}_{23} (\vec k_2 + \vec p) 
\mathcal E^{w}_{12} (\vec p) \\
&~~\equiv
\int_{\vec p}
\mathcal E^{nw, {\rm fin}}_{23} (\vec p) \Big(
\mathcal E^{nw, {\rm fin}}_{31} (\vec k_3 + \vec p) 
\mathcal E^{w}_{21} (\vec k_2 - \vec p) \Theta\lb |\vec k_3 + \vec p| - p\rb 
 - \mathcal E^{nw, {\rm fin}}_{31} (\vec k_3) 
\mathcal E^{w}_{21} (\vec k_2) \Big)
+ (\vec k_1 \leftrightarrow \vec k_2)\, .   \non
}
In the above regularization we used the fact that we can split the integral domain into two parts, each characterized by the poles at $\vec p \to -\vec k_2$ and $\vec p \to \vec k_1$, such that we can write
\eeq{
\int_{\vec p}\,
\mathcal E^{nw, {\rm fin}}_{13} (\vec k_1 - \vec p) 
\mathcal E^{nw, {\rm fin}}_{23} (\vec k_2 + \vec p) 
\mathcal E^{w}_{12} (\vec p) 
=  \int_{|\vec k_1 - \vec p| < |\vec k_2 + \vec p|} + \int_{|\vec k_1 - \vec p| > |\vec k_2 + \vec p|}\, .
}  
Furthermore, each contribution can be rewritten as 
\eq{
\int_{|\vec k_1 - \vec p| < |\vec k_2 + \vec p|}
\mathcal E^{nw, {\rm fin}}_{13} (\vec k_1 - \vec p) 
&\mathcal E^{nw, {\rm fin}}_{23} (\vec k_2 + \vec p) 
\mathcal E^{w}_{12} (\vec p) \non \\
& = \int_{p < |\vec k_3 + \vec p|}
\mathcal E^{nw, {\rm fin}}_{13} (\vec p) 
\mathcal E^{nw, {\rm fin}}_{32} (\vec k_3 + \vec p) 
\mathcal E^{w}_{12} (\vec k_1 - \vec p) \,.
}
At this point there are still remaining IR contributions arising from taking $\vec p \to \vec k_3$. We can further extract these IR modes by isolating the $p < k_3$ region from the integral above so that 
\eq{
\int_{\vec p, {\rm IR}}\,
\mathcal E^{nw,{\rm fin}}_{13} & (\vec k_1 - \vec p) 
\mathcal E^{nw,{\rm fin}}_{23} (\vec k_2 + \vec p) 
\mathcal E^{w}_{12} (\vec p) \non \\
&= 
\int_{\vec p}
\mathcal E^{nw, {\rm fin}}_{23} (\vec p) \Big(
\mathcal E^{nw, {\rm fin}}_{31} (\vec k_3 + \vec p) \Theta\lb |\vec k_3 + \vec p| - p\rb  - 
\mathcal E^{nw, {\rm fin}}_{31} (\vec k_3) \Big)
\mathcal E^{\rm w}_{21} (\vec k_2 - \vec p) \non\\
& ~~~~
+ \mathcal E^{nw, {\rm fin}}_{31} (\vec k_3) 
\int_{\vec p}
\mathcal E^{nw, {\rm fin}}_{23} (\vec p) \Big(
\mathcal E^{w}_{21} (\vec k_2 - \vec p) - 
\mathcal E^{w}_{21} (\vec k_2) \Big) + (\vec k_1 \leftrightarrow \vec k_2) \non \\
&\approx
\int_{\vec p}
\mathcal E^{nw, {\rm fin}}_{23} (\vec p) \Big(
\mathcal E^{nw, {\rm fin}}_{31} (\vec k_3 + \vec p) \Theta\lb |\vec k_3 + \vec p| - p\rb  - 
\mathcal E^{nw,{\rm fin}}_{31} (\vec k_3) \Big)
\mathcal E^{w}_{21} (\vec k_2 - \vec p) \non\\
& ~~~~
+ \mathcal E^{nw,{\rm  fin}}_{31} (\vec k_3) 
\lb e^{\half(\vec k_1 \cdot \vec k_2) \Sigma^2_{s}} - e^{- \half k^2_2 \Sigma^2_{s}} \rb 
 \frac{(\vec k_1 \cdot \vec k_2)}{k_2^2} P^w_{\rm L}(k_2) + (\vec k_1 \leftrightarrow \vec k_2)\, . 
 \label{eq:local_wiggle_approx}
}
In the second line above, we used the approximation that the wiggles are localized in the configuration space, corresponding to the configuration-space saddle point, i.e. 
\eq{
\int_{\vec p}
\mathcal E^{nw, {\rm fin}}_{23} (\vec p) 
\Big(
& \mathcal E^{w}_{21} (\vec k_2 - \vec p) - 
\mathcal E^{w}_{21} (\vec k_2) \Big) \non \\
&= e^{\frac{1}{2} \vk_{1,i} \vk_{2,j} A^{nw}_{ij}(\vec q_*) }  \tfrac{1}{2} \vk_{1,i} \vk_{2,j}   \int_{\vec q} \bq\  e^{-i \vec k_2 \cdot \vec q}  A^{w}_{ij}(\vec q)  
\int_{\vec p} \mathcal E^{nw, {\rm fin}}_{23} (\vec p) \lb e^{ i \vec p \cdot \vec q} - 1 \rb \non\\
&= e^{\frac{1}{2} \vk_{1,i} \vk_{2,j} A^{nw}_{ij}(\vec q_*) } \lb e^{\frac{1}{2} \vk_{2,i} \vk_{3,j}  A^{nw}_{ij}(\vec q_*)} - 1 \rb 
 \tfrac{1}{2} \vk_{1,i} \vk_{2,j}   \int d^3\bq\  e^{-i \vec k_2 \cdot \vec q} \ A^{w}_{ij}(\vec q) \non\\
&\approx - \lb e^{- \frac{1}{2} \vk_{2,i} \vk_{2,j}  A^{nw}_{ij}(\vec q_*)} - e^{\frac{1}{2} \vk_{1,i} \vk_{2,j} A^{\rm nw}_{ij}(\vec q_*)} \rb 
 \frac{(\vec k_1 \cdot \vec k_2)}{k_2^2} P_{\rm L}^{w}(k_2)  \non\\
&\approx \lb e^{\half (\vec k_1 \cdot \vec k_2) \Sigma^2_{s}} - e^{- \half k^2_2 \Sigma^2_{s}} \rb 
 \frac{(\vec k_1 \cdot \vec k_2)}{k_2^2} P_{L}^{w}(k_2)  \, .  
}
We see that we obtain two exponential contributions, one containing the angular dependence of the two modes $\vec k_1 \cdot \vec k_2$ and the second one depending only on $k_2^2$. One might wonder if the first piece can give rise to bispectrum configurations where the wiggles are enhanced. However, this does not happen as these contributions are exactly equal in magnitude but opposite in sign from the ones in the tree level part of Equation \ref{eq:B_wiggle_IR_safe_I}. Thus, combining all these results and cancelling all isolated IR contributions, we get
\eq{
B^{w}
&= - \Big( \mathcal E^{nw, {\rm fin}}_{32}(\vec k_3) (\vec k_1 \cdot \vec k_2)
+ \mathcal E^{nw,{\rm fin}}_{23}(\vec k_2 ) (\vec k_1 \cdot \vec k_3) \Big) e^{- \half k_1^2\Sigma^2_{s}} P_{\rm L}^{w}(k_1)/k_1^2
 \non\\
&~~~ +
\int_{\vec p}
\mathcal E^{nw,{\rm fin}}_{23} (\vec p) \Big(
\mathcal E^{nw,{\rm fin}}_{31} (\vec k_3 + \vec p) \Theta\lb |\vec k_3 + \vec p| - p\rb  - 
\mathcal E^{nw,{\rm fin}}_{31}(\vec k_3) \Big)  \mathcal E^{w}_{21} (\vec k_2 - \vec p) + (\vec k_1 \leftrightarrow \vec k_2) \non\\
&~~~ + \lb {\rm 2cycle} \rb \, .
\label{eqn:fs_saddle}
}
We have thus obtained the wiggle bispectrum in an explicitly IR-safe form, with the resummed long displacements acting on the linearized wiggle contribution. The mode coupling integral above can also be rewritten in terms of the $\vec k_1$ and $\vec k_2$ modes as 
\eq{
\int_{\vec p}
\mathcal E^{nw,{\rm fin}}_{23} (\vec p) \Big(
&\mathcal E^{nw,{\rm fin}}_{31} (\vec k_3 + \vec p) \Theta\lb |\vec k_3 + \vec p| - p\rb  - 
\mathcal E^{nw,{\rm fin}}_{31} (\vec k_3) \Big)
\mathcal E^{w}_{21} (\vec k_2 - \vec p) \\
&= \int_{\vec p}
\mathcal E^{nw,{\rm fin}}_{23} (\vec k_2 + \vec p) \Big(
\mathcal E^{nw,{\rm fin}}_{13} (\vec k_1 - \vec p) \Theta\lb | \vec k_1 - \vec p | - | \vec k_2 +\vec p | \rb  - 
\mathcal E^{nw,{\rm fin}}_{31} (\vec k_3) \Big)
\mathcal E^{\rm w}_{12} (\vec p) \, . \non
}

Let us consider the possible corrections to this IR resummed result. As stated above, we have linearized the result in $\PL^w$, and one can, of course, extend the analysis to include quadratic and higher-order contributions. Moreover, the localized-wiggle (saddle-point) approximation, used in Equation \ref{eq:local_wiggle_approx}, gives rise to correction terms discussed in Appendix \ref{app:saddle_point}. On top of these, however, we can also consider the residual contributions of the IR modes in the mode coupling integral in Equation \ref{eqn:fs_saddle}. Isolating the modes $p< \Lambda < k_{2,3}/2$ we can expand the integrand in the IR-safe mode coupling integral to obtain the correction due to these modes
%\eq{
%\mathcal E^{w}_{21} (\vec k_2)
%\int_{p<\Lambda}
%\mathcal E^{nw,{\rm fin}}_{23} (\vec p)&\Big(
%\mathcal E^{nw,{\rm fin}}_{31} (\vec k_3 + \vec p)  - 
%\mathcal E^{nw,{\rm fin}}_{31}(\vec k_3) \Big) \non\\
%& =\mathcal E^{w}_{21} (\vec k_2)
%\sum_{n=1}^\infty \frac{1}{n!}\int_{p<\Lambda}
%\mathcal E^{nw,{\rm fin}}_{23} (\vec p) (\vec p \cdot \nabla_{\vec k_3})^n 
%\mathcal E^{nw,{\rm fin}}_{31} (\vec k_3) \non\\
%& \sim \mathcal E^{w}_{21} (\vec k_2) \mathcal E^{nw,{\rm fin}}_{31} (\vec k_3)
%\sum_n\int_{p<\Lambda}
%\lb  \frac{p}{k_3}\rb^n \mathcal E^{nw,{\rm fin}}_{23} (\vec p) \,
%}
\begin{align}
    \int_{\vec p} \mathcal E^{nw,{\rm fin}}_{23} (\vec p) &\Big(\mathcal E^{nw,{\rm fin}}_{31} (\vec k_3 + \vec p) - \mathcal E^{nw,{\rm fin}}_{31} (\vec k_3) \Big) \mathcal E^{w}_{21} (\vec k_2 - \vec p) \non \\
    &= \sum_{n=1}^\infty \frac{1}{n!}  \nabla_{i_1 .. i_n}^n \mathcal E^{nw,{\rm fin}}_{31}(\bk_3) 
 \ \int_{\vec p} \bp_{i_1} ... \bp_{i_n} \mathcal E^{nw,{\rm fin}}_{23} (\vec p)\ \mathcal E^{w}_{21} (\vec k_2 - \vec p) \non \\
  &=  \sum_{n=1}^\infty \frac{1}{n!}  \nabla_{i_1 .. i_n}^n \mathcal E^{nw,{\rm fin}}_{31}(\bk_3)\ \int_{\vec q} e^{-i \bk_2 \cdot \bq}  \left(\frac{\partial^n \tilde{\mathcal{E}}_{23}^{nw, \rm fin}(\bq)}{\partial \bq_{i_1} ... \partial \bq_{i_n}} \right) \tilde{\mathcal E}^w_{12}(\bq) \non \\
  &\approx \sum_{n=1}^\infty \frac{1}{n!}  \nabla_{i_1 .. i_n}^n \mathcal E^{nw,{\rm fin}}_{31}(\bk_3)\ \int_{\vec q} e^{-i \bk_2 \cdot \bq}  \left(\frac{\partial^n \tilde{\mathcal{E}}_{23}^{nw, \rm fin}(\bq)}{\partial \bq_{i_1} ... \partial \bq_{i_n}} \right)_{q = r_s} \tilde{\mathcal E}^w_{12}(\bq)\, ,
\end{align}
where the $\nabla$ derivatives are with respect to $\bk_3$. Already in the second line we see that the smoothness of the no-wiggle component leads to a suppresssion, since the $\nabla$ derivatives multiplied by the integral over $\bp_i$ lead to a supppression of $(\Lambda / k_3)^n$ at each order in the Taylor series. However, the final correction is even further suppressed than this estimate, since by re-writing the correction as the Fourier transform in the final line we can see that the correction can again be evaluated using a a saddle-point approximation owing to the localization of the wiggles in configuration space. In Fourier-space this corresponds to the wiggle at $\bk_2 - \bp$ losing coherence with $\bk_2$ for $p$ wider than the wavelength of the oscillation. This leads to a correction of the form in Equation~\ref{eqn:ir_corrections} due to modes at the BAO scale, which is indeed not surprising since the two derivations use the same IR domain in different spaces. In the next subsection, where we explicitly show and compare various IR resummation schemes, we also explicitly evaluate the residual IR and UV corrections due to the mode coupling integral in Equation~\ref{eqn:fs_saddle}, showing that the latter are numerically highly suppressed as argued here.

\subsection{Numerical Comparison of Resummation Schemes}

\begin{figure}
    \centering
    \includegraphics[width=\textwidth]{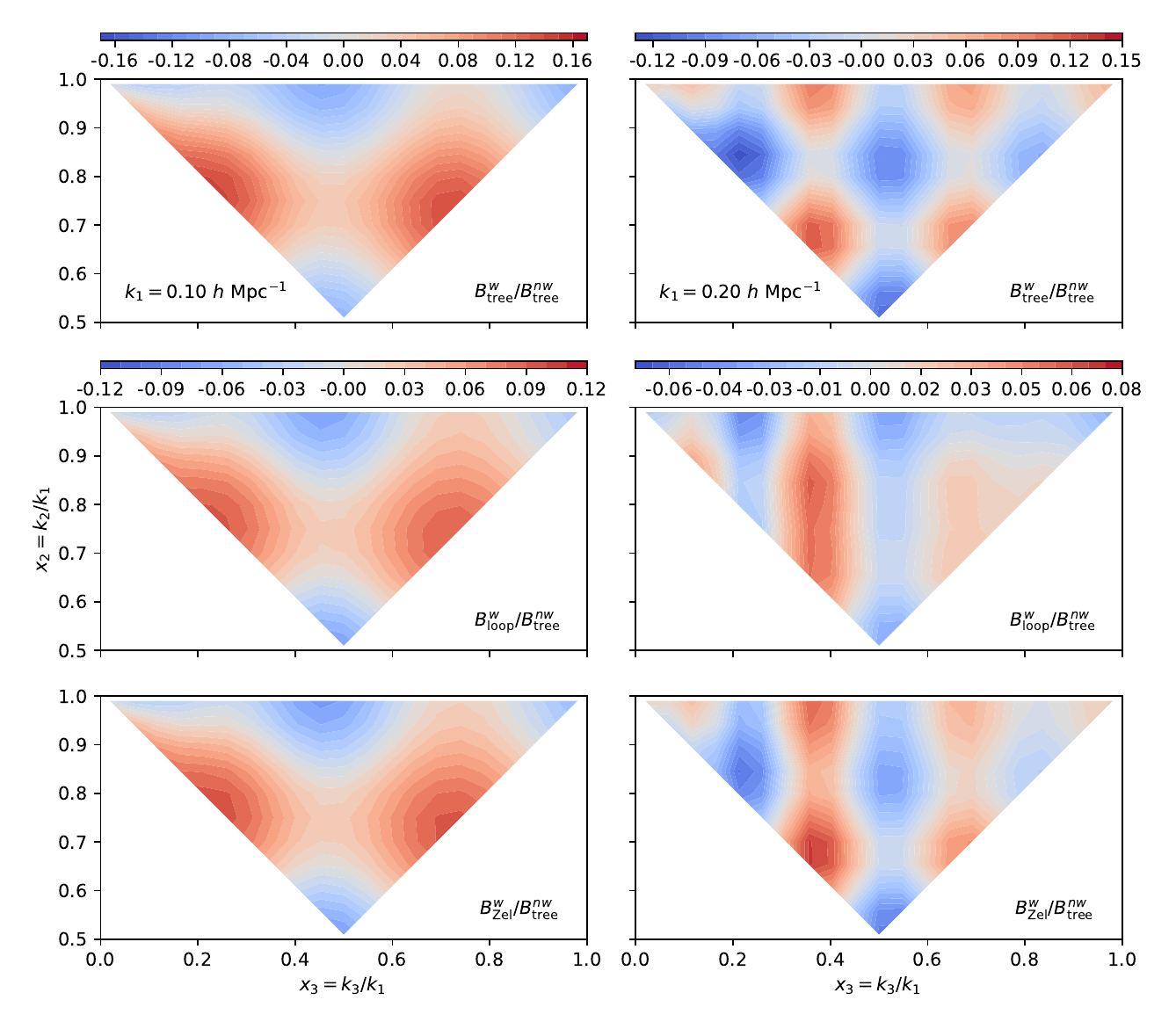}
    \caption{As in Figure~\ref{fig:triangle_plots} but with the BAO wiggle component isolated and normalized by the no-wiggle tree-level prediction. \edit{Deviations} in the wiggle phase and amplitude are evident in the $k = 0.20 \kMpc$ triangles\edit{---for example, the BAO feature in the upper right (equilateral) corner along the right triangle edge of the Zeldovich panel is in-phase, i.e. has the same sign (color) as the tree-level case, while the same region in the 1-loop panel has a flipped sign, mirroring the equilateral panel in Figure~\ref{fig:equilateral_squeezed_wiggles}}.}
    \label{fig:triangle_plots_wiggles}
\end{figure}

\begin{figure}
    \centering
    \includegraphics[width=0.9\textwidth]{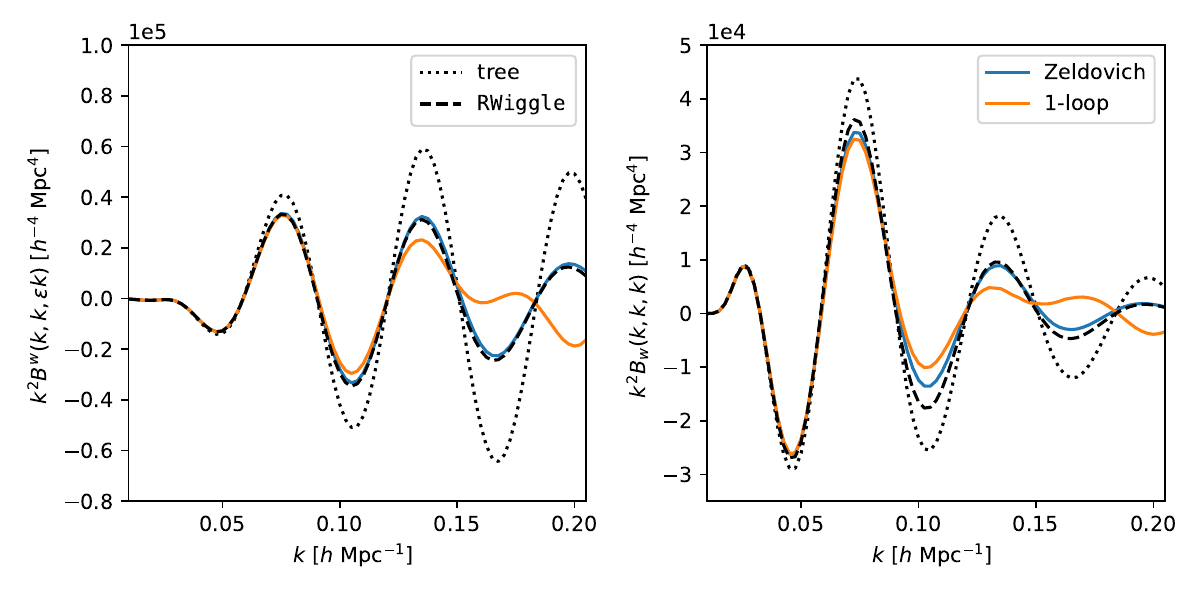}
    \caption{As in Figure~\ref{fig:numerical}, BAO in the squeezed (left) and equilateral (right) bispectrum configurations. The full Zeldovich calculation (blue) approaches the 1-loop (orange) and tree-level (dotted) predictions towards low $k$ but exhibit dramatically different oscillatory behavior towards smaller scales. The Zeldovich prediction is well approximated by an exponential damping of BAO wiggles in the linear power spectrum, particularly in the squeezed configuration.}
    \label{fig:equilateral_squeezed_wiggles}
\end{figure}

We can now numerically investigate the BAO signal in the bispectrum as predicted by various perturbation theory schemes at tree-level and 1-loop, pre- and post-IR resummation, using the full Zeldovich calculation as a benchmark where all the effects of long-wavelength linear displacements are manifestly taken into account. In order to do so it is convenient to extract the wiggle and no-wiggle components of bispectra as 
\begin{equation}
    B^w = B[\PL = \PL^{nw} + \PL^w] - B[\PL^{nw}],\ B^{nw}= B[\PL^{nw}]\, ,
\end{equation}
where the bispectra on the right-hand side are evaluated assuming the linear power spectra in the square brackets. \edit{This numerical definition includes contributions $\mathcal O(f_b^2)$ not considered in our derivations above whose interactions with nonlinearities should, however, be quite negligible.}

Figures~\ref{fig:triangle_plots_wiggles} and \ref{fig:equilateral_squeezed_wiggles} show the BAO components of the tree level, 1-loop and full Zeldovich bispectra, so defined, for bispectra of various side lengths and geometries. The 1-loop and full Zeldovich power spectra begin to behave quite differently at large wavenumbers, with the latter remaining mostly in phase with the tree-level predictions while the former veers off significantly and even changes sign.\footnote{\edit{We note that we have not adjusted for potential \textit{broadband} differences between the curves shown in Figure~\ref{fig:equilateral_squeezed_wiggles} or other similar plots in this paper, since in the Zeldovich case our $P_{nw}$ seems to do a sufficiently good job isolating broadband changes due to nonlinearities, and it is evident by eye that the difference between the 1-loop and Zeldovich curves in Figure~\ref{fig:equilateral_squeezed_wiggles} cannot be described by a smooth function of $k$.}} This is as expected because the damping effect by long displacements is controlled by a rather larger parameter $k^2 \Sigma^2_s$ and its effects become rather non-perturbative at the higher wavenumbers show in these plots. The enhancement of this large parameter at 1-loop in perturbation theory is described explicitly at the end of Appendix~\ref{app:ept_1loop}. 

In comparison to the Zeldovich 1-loop prediction, the dashed lines in Figure~\ref{fig:equilateral_squeezed_wiggles} show the prediction of the \textit{resummed} tree-level bispectrum using the \texttt{RWiggle} scheme. Evidently, this scheme captures the damping of the BAO component quite well, with particularly excellent agreement in the case of the squeezed configuration, though it slightly underpredicts the damping in the equilateral triangle. The agreement in the former case can be understood by noting that in the squeezed configuration the contribution due to the integral piece is quite small, so that the wiggles are essentially due to the $\mathcal{E} \mathcal{E}$ terms for which $\texttt{RWiggle}$ works almost perfectly.

\begin{figure}
    \centering
    \includegraphics[width=\textwidth]{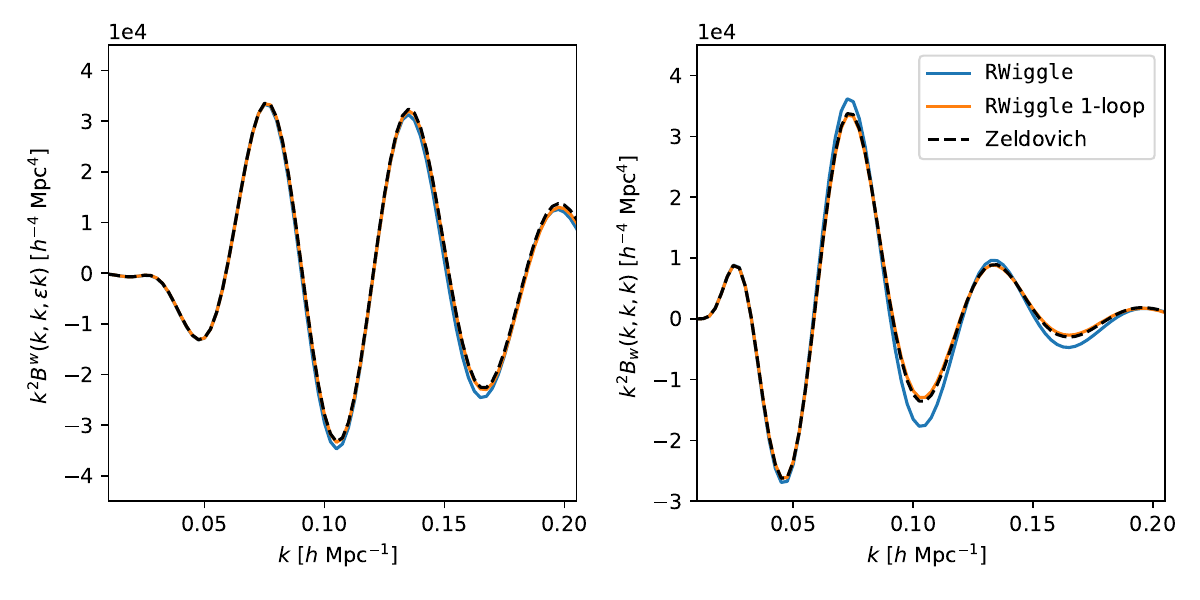}
    \caption{The 1-loop bispectrum resummed using the \texttt{RWiggle} scheme (orange), compared to the full Zeldovich bispectrum (black dashed) and the resummed tree-level bispectrum (blue). All three are in reasonably good agreement in squeeed triangles (left) other than a slight phase shift not captured by the tree-level spectrum, but the tree-level $\texttt{RWiggle}$ visibly underestimates the damping of the equilateral configuration even above $0.05 \kMpc$. }
    \label{fig:ept_ir_resummed}
\end{figure}

We can also compare the prediction of the Zeldovich calculation to the \texttt{RWiggle} at 1-loop order. In this case we adopt the scheme from ref.~\cite{Blas16} (see also \S\ref{sec:npoint})
\begin{equation}
    B^{\tt RWiggle}_{\rm 1-loop} = B_{\rm tree}\big[\PL^{nw} + (1 + \frac12 k^2 \Sigma_{s}^2)\ e^{-\frac12 k^2 \Sigma_{s}^2} \PL^w\big] + B_{\rm loop}[\PL^{nw} + e^{-\frac12 k^2 \Sigma_{s}^2} \PL^w] \, ,
\end{equation}
where the brackets indicate the bispectrum evaluated substituting the argument for the linear power spectrum. This comparison is shown in Figure~\ref{fig:ept_ir_resummed}: we see that the 1-loop corrections to $\texttt{RWiggle}$ almost entirely corrects for the discrepancy with the full Zeldovich calculation, including the small phase-shift visible in the squeezed configuration and the underpredicted damping in the equilateral one. This was not entirely unexpected; as shown in \S\ref{ssec:saddle} corrections to \texttt{RWiggle} are due to parametrically small density modes at the BAO scale, whose contributions are expected to be perturbative. 

\begin{figure}
    \centering
    \includegraphics[width=\textwidth]{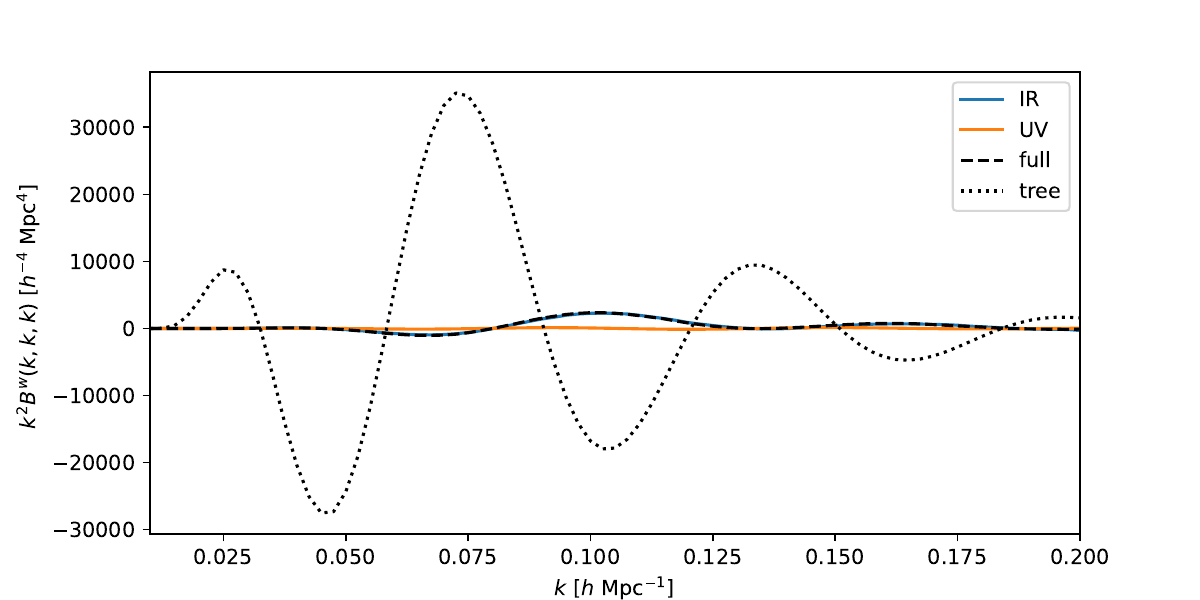}
    \caption{Contributions to the mode coupling integral when resumming the effects of linear displacements on the wiggle component in Equation~\ref{eqn:fs_saddle}. Contributions are split into IR and UV regions longer and shorter than the sides of the equilateral triangle, with the effect on the resummed wiggles almost entirely due to the IR modes.}
    \label{fig:ir_uv_split}
\end{figure}

To further check this intuition, Figure~\ref{fig:ir_uv_split} shows the correction to \texttt{RWiggle} derived in Equation~\ref{eqn:fs_saddle} split into contributions from large and small wavenumbers. Here we have evaluated the $\mathcal{E}$ in the mode-coupling integrals at tree-level for simplicity, since this captures the leading 1-loop corrections to the tree-level wiggles. The full mode-coupling integral is shown as a black-dashed line and is evidently composed almost entirely of the IR contribution (blue). The UV portion (orange), while nonzero, is close to an order of magnitude smaller than the IR one and out-of-phase with the tree-level oscillations.  This numerically validates the argument in \S\ref{sssec:config_saddle} and \ref{sssec:fs_saddle} that only specific IR modes close to the BAO scale contribute correcting the saddle-point approximation. Both contributions are in addition very small ($\approx$ 20x) compared to the tree-level prediction, showing that \texttt{RWiggle} captures the dominant effect of the linear displacements.

\subsection{Further IR-Resummation Schemes}

In addition to the direct Zeldovich calculation and \texttt{RWiggle} a few other schemes for IR resummation exist in the literature, particularly as applied to the power spectrum \cite{SenZal15,Schmittfull19}. The goal of this subsection is to briefly sketch their relation to our calculations and how they may be extended to the case of the bispectrum.

Ref.~\cite{SenZal15} constructed an IR resummation scheme of the two-point function with the goal of IR resummation operating on the usual EPT loop terms. This ensures that the IR resummed theory has approximately the same UV properties (counterterm structure and values) as the canonical, un-resummed, EPT. We follow the presentation from Appendix B of ref.~\cite{VlaWhiAvi15}, which re-derived this result starting from the canonical Zeldovich power spectrum.  
By defining  $A_{ij}(\vec q) = \la \Delta_i \Delta_j \ra = \Sigma^2 \delta^{\rm K}_{ij} - 2 \eta_{ij}(\vec q) $, we can write 
\eq{
\mathcal E^{\rm fin}_{12}(\vec p) 
&= \int_{\vec q}\, e^{i \vec p \cdot \vec q} \lb e^{\frac{1}{2} k_{1i} k_{2j}  A_{ij}(\vec q)} - e^{\frac{1}{2} k_{1i}k_{2j}  A_{ij}(\infty)}\rb  \non\\
&= \int_{\vec q}\, e^{i \vec p \cdot \vec q}\, e^{\frac{1}{2} k_{1i} k_{2j}  A_{ij}(\vec q)} \lb 1 - e^{k_{1i}k_{2j} \eta_{ij}}\rb  \non\\
&= - \sum_{n=1}^\infty \frac{1}{n!}  \int_{\vec q}\, e^{i \vec p \cdot \vec q}\, K_{12}(\vec q) \lb k_{1i}k_{2j} \eta_{ij} \rb^n \, ,
\label{eq:E_fin_expand_II}
}
where we introduce the Gaussian kernel 
$K_{12}(\vec q)  =  \exp\left\{\frac{1}{2} k_{1i} k_{2j}  A_{ij}(\vec q)\right\}$.
On the other hand, $\mathcal E^{\rm fin}$ can also be expanded in the conventional EPT fashion in powers of linear power spectra. This leads to the form 
\eq{
\mathcal E^{\rm fin}_{12} (\vec p)
&= \sum_{n=0}^\infty \mathcal E_{12}^{(n) - {\rm loop}}(\vec p) \, ,
}
where for $n=0$ we have the linear theory results, while the higher $n$ give us higher loop results
\eeq{
\mathcal E_{12}^{(n) - {\rm loop}}(\vec p) = 
\frac{1}{(n+1)!} \sum_{m=1}^{n+1} (-1)^m \binom{n+1}{m} \left\{ \lb \frac{1}{2}\lb \vk_1 \cdot \vk_2 \rb  \Sigma^2 \rb^{n-m+1} \big( k_{1i} k_{2j} \eta_{ij} \big)^{m}  \right\}_{\rm FT}  \, .
}
Up to one-loop order, this gives us
\eq{
\mathcal E_{12}^{\rm lin} &= - \Big\{ k_{1i} k_{2j} \eta_{ij}  \Big\}_{\rm FT} \, , \non\\
\mathcal E_{12}^{1- {\rm loop}} &= 
\frac{1}{2} \Big\{ \lb k_{1i} k_{2j} \eta_{ij} \rb^2  \Big\}_{\rm FT} -  \frac{1}{2} \big( \vk_1 \cdot \vk_2  \Sigma^2 \big) \Big\{ k_{1i} k_{2j} \eta_{ij}  \Big\}_{\rm FT} \, .
}
where in the one-loop term we recognise the familiar $P_{22}+P_{13}$ structure. Also, note that these are different from the usual power spectrum loops (in the Zeldovich approximation) since they depend on three wave vectors $\vec k_1$, $\vec k_2$ and $\vec p$. However, aligning these as $\vec k_1=-\vec k_2=\vec p$ reproduces the power spectrum results, as was noted below Equation \ref{eqn:epsilon}.

Utilising this EPT style expansion, it can be shown that Equation \ref{eq:E_fin_expand_II} can be rewritten as 
\eq{
- \sum_{n=1}^\infty \frac{1}{n!} \lb k_{1i}k_{2j} \eta_{ij} \rb^n
= \sum_{n=0}^\infty \sum_{m=0}^n \left[ K^{-1}_{12} \right]_{n-m} \left \{\mathcal E_{12}^{(m) - {\rm loop}}\right\}_{\rm FT}
}
where we have introduced the inverse of the $K$ kernel and its expansion
\eq{
K^{-1}_{12}(\vec q)  &= \exp\left\{ - \frac{1}{2} k_{1i} k_{2j}  A_{ij}(\vec q)\right\} \, , \non\\ 
\left[K^{-1}_{12} \right]_{n}(\vec q)  &= \frac{1}{n!} \lb  k_{1i}k_{2j} \eta_{ij}  - \frac{1}{2} (\vk_1 \cdot \vk_2)\Sigma^2 \rb^n \, .
}
Combining these results, we can write 
\eq{
\mathcal E^{\rm fin}_{12}(\vec p) 
&= 
\sum_{n=0}^\infty \sum_{m=0}^n
\int_{\vec p'} \mathcal M^{(n-m)}_{12} \lb {\vec p - \vec p'}\rb \mathcal E_{12}^{(m) - {\rm loop}}(\vec p') \, ,
}
where the momentum space kernel $\mathcal M$ contains only effects of the long displacements
\eeq{
\mathcal M^{(n)}_{12}(\vec k) = \int_{\vec q}\, e^{i \vec k \cdot \vec q}\, K_{12}(\vec q) \left[K^{-1}_{12} \right]_{n}(\vec q) \, .
}
Note that, strictly speaking, even in the ordinary power spectrum case, the integral above is not a simple convolution of $\mathcal M^{(n)}$ kernels with the EPT loops, given that there are always external modes present in the loops of $\mathcal E_{12}$. Nonetheless, it is easy to see that this result reduces to EPT in the case that $\mathcal M^{(0)}$ is a delta function and the rest of $\mathcal M^{(m)}$ vanish. Moreover, ref.~\cite{SenZal15} shows that a good approximation of this result can also be achieved when the external $\bk_{1,2,3}$ and internal $\bp$ modes in $\mathcal E_{12}$ are identified, and thus the latter term indeed corresponds to the EPT result. This analysis can be extended to the bispectrum, for example, using Equation~\ref{eqn:epsilon_to_b}, or by generalizing $K_{12}(\bq)$ to the bispectrum exponent in Equation~\ref{eqn:zel_exponent}. 

Another related method to resum Zeldovich displacements was described in ref.~\cite{Schmittfull19} using the so-called \textit{shifted operators}
\begin{equation}
    \tilde{O}(k) = \int d^3\bq \ e^{-i \bk \cdot(\bq + \Psi(\bq))}\ O(\bq).
\end{equation}
In this scheme the matter clustering ($F(\bq) = 1$) term in LPT (Eqn.~\ref{eqn:bispectrum}) is re-written in terms of bias operators, such that for example the Zeldovich matter density is given by \cite{Schmittfull19}
\begin{align}
    \delta_{\rm Zel}(\bk) &=  \int d^3\bq \ e^{-i \bk \cdot(\bq + \Psi)} \Big\{ \delta(\bq) + \half \big( s^2(\bq) - \frac23 \delta^2(\bq) \Big)  + \ldots \big\}  \equiv \tilde{\delta}(\bk) + \half \tilde{\mathcal G}_2(\bk) + \ldots
\end{align}
up to quadratic order\edit{, where $s^2$ is the square of the traceless shear field}. The shifted operator basis is an intermediate between Lagrangian and Eulerian perturbation theory, almost keeping the form of the latter but with bulk displacements kept to arbitrary order in the exponent rather than expanded perturbatively. A useful fact is that \edit{both the nonlinear matter and galaxy density fields can be written in terms of the shifted operators above with varying coefficients and, conveniently, the bispectrum for the shifted operators up to the order shown above can be expressed in terms of generalizations of $\mathcal{E}$, as we describe in more detail in Appendix~\ref{app:biased_tracers}. While the numerical evaluation of the two alternative schemes discussed above is beyond the scope of this work, their mathematical structure is rather similar to the direct Zeldovich calculation presented in this work, and we outline the steps towards these extensions in the aforementioned Appendix.}

\section{IR Resummation for N-point Functions}
\label{sec:npoint}

We can also extend our analysis to the IR resummation of n-point functions beyond the bispectrum. To do so, we need to consider the generalised version of the cumulant in Equation~\ref{eqn:zel_exponent}
\begin{equation}
    \avg{e^{-i \sum_{n=1}^N k_{n,i} \Psi^{(1)}_i(\bq_i)}}_{\bk_{\rm tot} = 0} = \exp\left\{ \half \sum_{n=1}^{N} \sum_{m > n} k_{n,i} k_{m,j} A_{ij}(\bq_{nm}) \right\}\, .
\end{equation}
Here, the sum runs over all \textit{distinct} pairs $(n,m)$, and we have used the identity Equation~\ref{eqn:DeltaiNDeltajN}. In the above we have defined $\bq_{nm} = \bq_n - \bq_m$. As before it is useful to shift to a coordinate system $\bq_{i<N} \rightarrow \bq_{i<N} - \textbf{Q}$, with  $\textbf{Q} = \bq_N$, such that the expectation values can be written independently of $\textbf{Q}$ by translation invariance. The Fourier-space N-point function is then given by
\begin{equation}
    C_N(\bk_1, ..., \bk_{N-1}) = \left( \prod_{a = 1}^{N-1} \int_{\bq_a} e^{-i\bk_a \cdot \bq_a }\right)\ \exp\left\{ \half \sum_{n\neq m} k_{n,i} k_{m,j} A_{ij}(\bq_{nm}) \right\}\ f_N(\bq_1, ..., \bq_{N-1})\, ,
\end{equation}
where we have used ``$n \neq m$'' as the shorthand for the sum over distinct pairs. Here, $f_N$ captures terms due to galaxy bias and higher-order displacements and is equal to unity for matter in the Zeldovich approximation. We can equivalently express the above in terms of $\mathcal{E}$ as
\begin{equation}
    C(\bk_1, ... , \bk_{N-1}) = \prod_{n \neq m } \int_{\bp_{nm}} \ \mathcal{E}(\bk_n, \bk_m, \bp_{nm}) \prod_{a=1}^{N-1} (2\pi)^3 \delta_D\big(\bk_a - \sum_{n > a} p_{an} + \sum_{n<a} p_{na}\big)\, ,
\end{equation}
where the momentum conservation comes from integrating over the vertices $\bq_a$, and we remind the reader that there is no sum over the N$^{\rm th}$ vertex. Note that the bispectrum and power spectrum are special cases in that there is only one fewer vertex (2 and 1) than pairwise separations $\bq_{nm}$ (3 and 2), leading to a reduction in the number of integrals in Fourier space. With the trispectrum, for example, one has to do an equal number of integrals in each space. For higher N-point functions there is a rapid scaling in the number of internal momenta $\bp_{nm}$ with $N$, though the Fourier space integral still has the advantage of being smooth compared to the Fourier transforms required in configuration space.

Separating the contributions to the above integral via a wiggle-no wiggle split it is easy to work out the effect of long-wavelenght displacements from on BAO wiggles in the N-point functions. In particular, expanding the wiggles to linear order we will have contributions that are, schematically,
\begin{align}
   C_w \sim \left( \prod_{m = 2}^{N-1} \int_{\bq_m} e^{-i\bk_m \cdot \bq_m} \right) &\exp\left\{ \half \sum_{ n,m > 1}^{n \neq m} k_{n,i} k_{m,j} A_{ij}(\bq_{nm}) \right\} \nonumber \\
   &\int_{\bq_1} e^{-i\bk_1 \cdot \bq_1}   \exp\left\{ \half \sum_{m\neq 1} k_{1,i} k_{m,j} A_{ij}(\bq_{1m}) \right\}  \xi_w(\bq_{1})\, .
   \label{eqn:c_w}
\end{align}
Here we have chosen a BAO features in the $\bq_{1} = \bq_{1N}$ leg of the N-point function $f_N$ in Lagrangian space but, since the final expression is invariant under permutations of the vertices we can do so without loss of generality. The top line of Equation~\ref{eqn:c_w} is smooth in $\bq_1$ so we can ignore it. If we make the same approximation $\bq_{1m} \rightarrow \bq$ as in the bispectrum in the previous section then the exponent in the second line becomes at the saddle point
\begin{equation}
    \half k_{1,i} \left( \sum_{m\neq n} k_{m,j} \right) A_{ij}(\bq_{s}) = -\half k_1^2 \Sigma^2_{s} \, ,
\end{equation}
by momentum conservation. This implies that the effects of long-wavelength displacements on wiggles in N-point functions can be well-described, at leading order in the wiggles, by the substitution
\begin{equation}
    P_w(k) \rightarrow e^{-\half k^2 \Sigma^2_{s}} P_w(k)\, .
\end{equation}
This prescription is a consequence of translation invariance and momentum conservation; it is interesting to observe that this condition is also satisfied for N-point functions with collapsed legs, such that the loops can be computed with this substitution as well to approximate the effects of IR displacements.

\section{Conclusions}
\label{sec:conclusions}

\begin{figure}
    \centering
    \includegraphics[width=\textwidth]{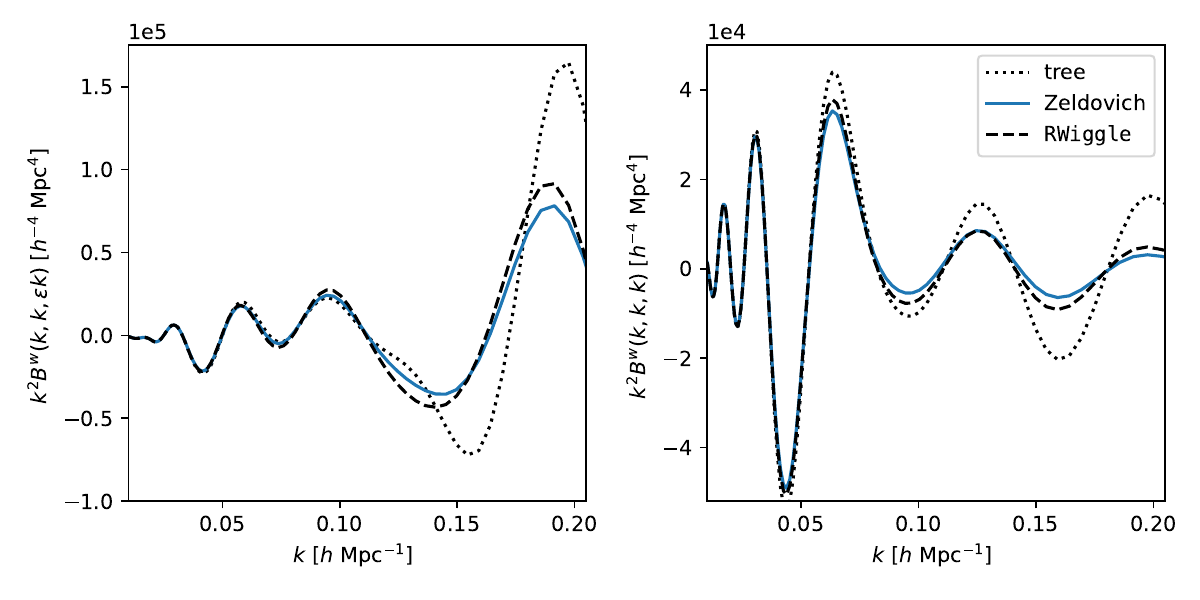}
    \caption{Squeezed (left) and equilateral (right) Zeldovich bispectra when the linear power spectrum has logarithmic wiggles on top of our fiducial $\Lambda$CDM one. The dotted and dashed lines show the tree level and $\texttt{RWiggle}$ predictions---the latter captures some of the damping due to the linear displacements, as in the case of the BAO, but with still clearly visible differences in the amplitude.}
    \label{fig:logwiggles}
\end{figure}

The presence of sharp, localized features in the linear correlation function---or, oscillatory ``wiggles'' in the power spectrum---lead large-scale structure observables to be sensitive to the bulk motion of matter on the large-scales at which these features reside. In $\Lambda$CDM universes close to our own, the contributions of these infrared (IR) modes to galaxy N-point statistics can be numerically quite large, such that they need to be resummed in order for perturbative calculations to be well-behaved. Indeed, this is the case both for features that are the bread and butter of modern galaxy surveys, i.e. the baryon acoustic oscillations standard ruler, and proposed targets of future experiments, e.g. features in the primordial power spectrum that probe the particular shape of the inflaton potential. A natural way to formulate this procedure, known as IR resummation, is within the Lagrangian picture of fluid mechanics, wherein fluids are described via the displacements of individual fluid elements which give rise to these bulk flows on large scales. Within Lagrangian perturbation theory (LPT), the effect of the bulk displacements can be understood as the smearing of the correlation function due to \textit{pairwise} displacements of fluid elements separated by the scale of the feature (e.g. the BAO radius $r_d$).

In this paper, we build on previous works on the galaxy 2-point function to study the galaxy 3-point function, or bispectrum, in Lagrangian perturbation theory. In the first part, we extend past techniques used to efficiently compute the resummed LPT power spectrum to the resummed bispectrum, focusing specifically on the case of the bispectrum within the Zeldovich approximation which contains all the long-wavelength displacements we wish to include. Our method involves computing the Fourier transform of the variance of the pairwise displacement, projected along arbitrary wavenumbers, and convolving the resulting kernels in Fourier space. These kernels contain an IR pole when the wavenumber $\bp = \textbf{0}$ whose effects cancel in the full calculation, and in order to properly treat their convolution numerically we split the integral into Voronoi regions closest to each pole and re-organize the expression for the integrand in each such that IR divergences explicitly vanish. This is a generalization of the so-called IR-safe integrands in Eulerian perturbation theory, with the effects of bulk displacements now carried to arbitrary order. We make the code to perform these calculations, \texttt{triceratops}, publicly available.

In the second part of the paper, we focus on computing the effect of the resummed displacements on features in the linear 2-point function, taking the BAO peak as a particular example. In the 2-point function, the presence of such a feature allows for a simple saddle-point approximation of these displacements at the peak, leading to a simple Gaussian damping of the power spectrum wiggles in Fourier space. The situation in the 3-point function is not so simple: here, requiring that two points in a triangle be separated by the BAO radius nonetheless leaves substantial freedom for the remaining point and two legs of the triangle to move around. However, we can nonetheless perform a saddle point approximation of the Zeldovich displacements in the bispectrum---we show that this leads to the familiar form where the effect of the displacements is a Gaussian smoothing of the linear power spectrum in each leg $k$, which we call \texttt{RWiggle}, with corrections from the extra degrees of freedom due to perturbatively small density modes on the BAO scale, suppressed by inverse powers of $k r_d$. This conclusion holds for arbitrary $N$-point functions due to momentum conservation. We numerically compare the wiggle component computed in this approximation to that in the full Lagrangian calculation, finding close agreement, especially at 1-loop order after the IR resummation of the wiggle component. We also make contact with other IR resummation schemes in the literature, deriving an extension of the scheme developed for the power spectrum in ref.~\cite{SenZal15} from the Lagrangian calculation and showing how the bispctrum in the \textit{shifted-operator} formalism developed for field-level analyses in ref.~\cite{Schmittfull19} can be computed uinsg the techniques in this paper.

We have thus established that resummed 1-loop perturbation theory very-well captures the effect of long-wavelength displacements on the BAO signal in cosmological N-point functions. However, it is worth noting that the direct Lagrangian treatment has distinct advantages over the \texttt{RWiggle} prescription, in that the wiggle (feature) component does not have to be distinctly treated. One advantage of this is that in the Lagrangian method one does not need to perform the wiggle/no-wiggle split, which is somewhat arbitrary, since there is no clear criterion that the smooth and wiggle components must satisfy, and indeed likely need to be adapted individually for each inflationary model treated. Moreover, features in the linear power spectrum beyond the BAO, e.g. inflationary signatures in the primordial spectrum, may have nonlinear dispersion relations not localized to any particular scale in configuration space---such a case is handled out-of-the-box in LPT but needs to be handled on a case-by-case basis in \texttt{RWiggle}, since each frequency in Fourier space corresponds to a distinct IR scale. This makes the Lagrangian treatment indispensable for future spectroscopic searches for exotic physics.

As a simple example let us consider oscillations in the bispectrum due to logarithmic oscillations in the primordial power spectrum
\begin{equation}
    \PL(k) = \PL^{\Lambda \rm CDM}(k) \left(1 + A \sin(\omega \log\frac{k}{k_\ast})\ e^{-(k r_d)^2/2}\right) \, ,
\end{equation}
which can appear in certain models of inflation \cite{Flaugher10,Slosar19}. Previous works \cite{Vasudevan19,ChenVlahWhite20b} have shown that, in the limit of large log frequency, these logarithmic oscillations are approximately damped in a wavenumber-dependent fashion\footnote{In LPT this corresponds to the limit in which the oscillations can be locally well-approximated by a Taylor expansion
\begin{equation}
    \phi(k) = \omega f(k) = \omega f(k_0) + \omega f'(k_0) (k - k_0) + ...
\end{equation}
Around $k_0$ the oscillations correspond to a physical scale of $q(k) = \omega f'(k_0) = \omega / k_0$ for logarithmic oscillations, though we note that this analysis in principle applies to any nonlinear dispersions.
} by $\Sigma^2(k) = X(q(k)) + Y(q(k))$, where $q(k) = \omega/k$ while also exactly computed in the full LPT calculation of the power spectrum. Using the results in this paper, we can straightforwardly extend these calculations to the bispectrum, as we have done in Figure~\ref{fig:logwiggles}. Here we have followed ref.~\cite{ChenVlahWhite20b} and chosen $A = 0.05$, $\omega = 10$, $k_\ast = 0.05 \kMpc$ and $r_d = 2.5\ h^{-1} \text{Mpc}$. As in the case of the BAO the \texttt{RWiggle} prescription at tree level qualitatively describes the damping of the wiggles, though unlike there deviations can be seen even in the squeezed configuration. For current and future surveys, the galaxy bispectrum carries substantial promise as a probe of primordial features on a comparable footing with the power spectrum \cite{Ballardini24}; however, since searches for primordial features need to recover or constrain not just the frequency but the \textit{amplitude} of power spectrum oscillations it will be important to consider these differences in IR resummation schemes when using the bispectrum as a probe of inflationary features. For that, a full treatment of the bispectrum in Lagrangian perturbation theory beyond what was developed here, or Eulerian calculations at 1-loop order, will be essential.

Indeed, while we have focused mostly on the matter power spectrum in the Zeldovich approximation in this work, our calculations have immediate implications for galaxy n-point functions in real and redshift space. Extending to redshift-space is a simple matter of coordinate transformations. The linear displacements are converted to their redshfit-space counterparts via a matrix transformation boosting the line-of-sight ($\hat{n}$) components, $\Psi^{(1)}_{s,i} = R_{ij} \Psi^{(1)}_j$, where $R_{ij} = \delta_{ij} + f(z) \hat{n}_i \hat{n}_j$ where $f(z)$ is the linear growth rate. This is equivalent to a remapping of the wavenumbers, since displacements always appear in the combination $k_i \Psi_{s,i} = (R. \textbf{k})_i \Psi_i$. In \texttt{RWiggle} this is equivalent to setting $k^2 \Sigma_s^2 \rightarrow k^2 ( 1 + f(2+f)\mu^2) \Sigma_s^2$, where $\mu = \hat{n} \cdot \hk$.  The extension to biased tracers can be similarly accomplished by letting the bias functionals in Equation~\ref{eqn:bispectrum} depend on local observables like the tidal tensor. Since this extension is not expected to change any conclusions about the role of infrared displacements, which instead live in the exponent of Equation~\ref{eqn:bispectrum}, we will leave detailed calculations to future work. The interested reader is, however, directed to Appendix~\ref{app:biased_tracers} for expressions in real-space up to quadratic order in the bias expansion.

\acknowledgments

SC would like to thank the CECC in Taiwan, which generously hosted him in the early stages of this project.
SC acknowledges the support of the National Science Foundation at the Institute for Advanced Study.
Z.V. acknowledges the support of the Kavli Foundation.
MW is supported by NASA and the DOE.

This research has made use of NASA's Astrophysics Data System and the arXiv preprint server.
This research is supported by the Director, Office of Science, Office of High Energy Physics of the U.S. Department of Energy under Contract No. DE-AC02-05CH11231, and by the National Energy Research Scientific Computing Center, a DOE Office of Science User Facility under the same contract.
 
\appendix

\section{Numerical Evaluation of $\mathcal{E}$}
\label{app:epsilon_numerics}
In \S\ref{sec:direct_integration} we introduced a generalized version of the Zeldovich power spectrum, the function $\mathcal{E}(\bk_1, \bk_2, \bp)$, through which the effects of large linear displacements can be captured. The goal in this appendix is to develop numerical techniques to efficiently evaluate this function. In \S\ref{ssec:fiducial_method} we outline one way to perform this calculation by performing an angular decomposition of the integral into parts that can be rapidly computed by Hankel transform. \S\ref{ssec:linear_theory} shows how the first few terms of this decomposition reduce to linear theory at leading order. Finally, \S\ref{ssec:alternative} presents an alternative method to perform this calculation whose form will also be useful in computing the angular structure of the BAO resummation in \S\ref{sec:derivation_xpy}.

\subsection{Angular Decomposition and Fast Evaluation Using Hankel Transforms}
\label{ssec:fiducial_method}

Let us now examine $\mathcal{E}$ in more detail. By symmetry, we have the angular dependence
\eeq{
\mathcal E(\vec k_1, \vec k_2, \vec p) 
= \mathcal E\lb k_1, k_2, p, \hat k_1 \cdot \hat k_2, \hat k_1 \cdot \hat p, \hat k_2 \cdot \hat p \rb
= \int d^3q\, e^{-i \vec p \cdot \vec q} e^{+ \frac{1}{2} \vk_{1,i} \vk_{2,j}  A_{ij}(\vec q)}\, , 
}
where $A_{ij} = X\df^K_{ij} + Y \hat q_i \hat q_j$ \cite{CLPT}.  
This gives us 
\eeq{
\frac{1}{2} \vk_{1,i} \vk_{2,j}  A_{ij}
= \frac{1}{2} k_1 k_2 \left[ \hat k_1 \cdot \hat k_2 X(q) +  (\hat k_1 \cdot \hat q) (\hat k_2 \cdot \hat q) Y(q) \right]\, .
}
and thus 
\eq{
\mathcal E & \lb k_1, k_2, p, \hat k_1 \cdot \hat k_2, \hat k_1 \cdot \hat p, \hat k_2 \cdot \hat p \rb \non\\
&\hspace{2.5cm} = \int q^2dq\,  e^{\frac{1}{2} (\vec k_1 \cdot \vec k_2) \left[ X(q) + Y(q) \right] }
\mathcal I \lb qp, \tfrac{1}{2} k_1k_2 Y(q)  , \hat k_1 \cdot \hat k_2 , \hat k_1 \cdot \hat p, \hat k_2 \cdot \hat p \rb \, .
}
We are thus faced with solving integrals of type 
\eeq{
\mathcal I \lb \alpha, \beta, \hat k_1 \cdot \hat k_2, \hat k_1 \cdot \hat p, \hat k_2 \cdot \hat p  \rb
= \int d\Omega_{\hat q}\,
e^{-i \alpha \hat p \cdot \hat q + \beta \hat k_{1,i} \hat k_{2,j} \lb \hat q_i \hat q_j - \delta^K_{ij} \rb}
= \sum_{n=0}^\infty \frac{\beta^n}{n!} I_n\, .
}
The key part to compute is the integral
\eeq{
I_n = \int d\Omega_{\hat q}\,
\left[ \hat k_{1,i} \hat k_{2,j} \lb \hat q_i \hat q_j - \delta^{\rm K}_{ij} \rb \right]^n
e^{i \alpha \hat p \cdot \hat q} \, .
}
We can be a bit more general by replacing $\hat p$ with $\vec p$. We thus have
\eeq{
I_n = \int d\Omega_{\hat q}\,
\left[ \hat k_{1,i} \hat k_{2,j} \lb \hat q_i \hat q_j - \delta^{\rm K}_{ij} \rb \right]^n
e^{i \alpha \vec p \cdot \hat q} \, .
}
Notice that these integrals form the 
recursion 
\eeq{
I_{n+1} 
= -  \hat k_{1,i} \hat k_{2,j}  \lb \frac{1}{\alpha^2} \frac{\partial^2}{\partial p_i\partial p_j} +  \delta^{\rm K}_{ij} \rb I_{n}\, .
}
For $n=0$ we have
\eq{
I_{0} = 4\pi j_0(\alpha p)\, ,
}
and thus 
\eeq{
I_{n} = 4\pi \bigg[ - \hat k_{1,i} \hat k_{2,j}  \lb \frac{1}{\alpha^2} \frac{\partial^2}{\partial p_i\partial p_j} +  \delta^{\rm K}_{ij} \rb \bigg]^n j_0(\alpha p)\, .
}
We note that in case when $\hat k_1 = -\hat k_2 = \hat p$ we have
\eeq{
I_n = 4\pi \Big( 1 + \partial_{\alpha p}^2 \Big)^n j_0(\alpha p)\, = 
4\pi ~ n! \lb \frac{2}{\alpha p} \rb^n j_n(\alpha p)\, ,
}
recovering the results for the power spectrum in ref.~\cite{VlaSelBal15}.

In order to find a closed expression of the solution we try the ansatz
\eeq{
I_n =  (-1)^n 4\pi \lb F_n (a, b, c, \alpha p)  \frac{j_n(\alpha p)}{(\alpha p)^n} + (\alpha p)^2 G_n(a, b, c, \alpha p) \frac{j_{n+1}(\alpha p)}{(\alpha p)^{n+1}} \rb \, ,
}
with $a= \hat k_1 \cdot \hat k_2$, $b= \hat k_2 \cdot \hat p$ and $c= \hat k_1 \cdot \hat p$.
This gives us the following recursive solutions
\eq{
F_{n+1}&= \hat k_{1,i} \hat k_{2,j} \Big( \lb 2 G_{n,j} - F_{n,j} \rb p_i + (i \leftrightarrow j) + (2n+3) \tfrac{F_{n,ij}}{\alpha^2} + (\alpha p)^2 \tfrac{G_{n,ij}}{\alpha^2} \Big) \\
&\hspace{6.35cm} 
+ 2 a (n+1) F_n + \lb 2 a  - (\alpha p)^2 (b c -a) \rb G_n   \, ,\non\\
G_{n+1}&=  \hat k_{1,i} \hat k_{2,j} \Big( - \tfrac{F_{n,ij}}{\alpha^2} - G_{n,j} p_i + (i \leftrightarrow j) \Big)  + (b c - a) F_n  + \big( (2n+1) b c - a \big) G_n \, , \non
}
where ${}_{,i}\equiv \partial/\partial p_i$. Given that $F_0 = 1$ and $G_0=0$ we can run the recursion and get for the first few orders 
\eq{
F_{1}&= 2a \, ,\\
G_{1}&= b c - a \, , \non
}
and 
\eq{
F_{2}&= 1 + 7a^2 - (\alpha p)^2 (b c -a)^2   \, ,\\
G_{2}&= - a^2 - b^2 - c^2 - 4 a b c + 7 b^2 c^2  \,. \non
}
This motivates the further ansatz
\eq{
F_n (a, b, c, z) &= \sum_{m=0}^{\lfloor n/2 \rfloor} (-1)^{m} f_{n,m}(a,b,c) z^{2m} \, , \\ 
G_n (a, b, c, z) &= \sum_{m=0}^{\lfloor (n-1)/2 \rfloor} (-1)^{m} g_{n,m}(a,b,c) z^{2m} \, .  \non
}
Introducing operators
\eq{
\hat O_{1}(m)&=
\lb a - bc \rb \partial_b
+ \lb 1 - c^2 \rb  \partial_c
+ 2m c \, , \\
\hat O_{2}(m)&=
\lb 1 - b^2 \rb \partial_b 
+ \lb a - bc \rb  \partial_c 
+ 2m b \, , \non\\
\hat O_{12}(m) &=
 \lb a - bc \rb \lb 1 - b^2 \rb \partial^2_b 
 + \lb a - bc \rb \lb 1 - c^2 \rb \partial^2_c 
 +  \lb  \lb 1 - c^2 \rb \lb 1 - b^2 \rb  +  \lb a - bc \rb^2 \rb  \partial_b \partial_c \non\\
&~~ + \lb 2m (c + a b - 2b^2 c) + 3 b^2 c - 2ab - c \rb \partial_b
 + \lb 2m (b + a c - 2b c^2) + 3 b c^2 - 2ac - b \rb \partial_c \non\\
&~~ + 2m \big( a + 2(m-1)c b \big)  \, , \non
}
and since $\hat O_{12}(0) \tilde f_{n,0} = 0$ we end up with recursion relations 
\eq{
f_{n+1,m} &= \lb 2 c \hat O_{2}(m) + 2 b \hat O_{1}(m) + \hat O_{12}(m) + 2 a \rb g_{n,m} \\
&~~~ + \lb - c \hat O_{2}(m) - b \hat O_{1}(m) + 2 a (n+1) \rb f_{n,m} - (2n+3) \hat O_{12}(m+1) f_{n,m+1}  + (b c - a ) g_{n,m-1} \, , \non\\
g_{n+1,m} &= \hat O_{12}(m+1) f_{n,m+1} + (b c - a) f_{n,m} 
+\Big( - c \hat O_{2}(m) - b \hat O_{1}(m) + (2n+1) b c - a \Big) g_{n,m} \,. \non
}
These reproduce the same results as obtained before.
One can imagine pushing the recursion one layer further, assuming a polynomial form for $f_{n,m}(a,b,c)$ and $g_{n,m}(a,b,c)$. However, this does not seem practical, given that the resulting expression for coefficients requires three additional indices for powers of $a$, $b$ and $c$.

We can use the form above to factor out the $p$ dependence of these integrals.
\begin{align}
    \mathcal E &= 4\pi \sum_n \frac{(-1)^n}{n!}\int dq\ q^2 \ e^{\frac{1}{2}(\bk_1\cdot\bk_2)(X+Y)} \beta^n  (pq)^{-n} \Big ( F_n - (pq)^{2} \big( \frac{n\Theta_{n>0}}{\beta} \big) G_{n-1}  \Big ) j_n(pq) \nonumber \\
    &= \sum_{n,m} 4\pi \frac{(-1)^{n+m}}{n!}\int dq\ q^2 \ e^{\frac{1}{2}(\bk_1\cdot\bk_2)(X+Y)} \beta^n  (pq)^{2m-n} \Big ( f_{n,m} + \big( \frac{n\Theta_{n>0,m>0}}{\beta} \big) g_{n-1,m-1} \Big ) j_n(pq) \nonumber \\
    &\equiv \sum_{n,m} \frac{(-1)^{n+m}}{n!} p^{2m-n} \epsilon_{n,m}(p)
\end{align}
where as a reminder $F_n$ and $G_n$ are functions of $a, b, c$ and $pq$. In the above we have defined
\begin{equation}
    \epsilon_{n,m}(p) = 4\pi  \int dq\ q^2 \ e^{\frac{1}{2}(\bk_1\cdot\bk_2)(X+Y)} \beta^n  q^{2m-n} \Big ( f_{n,m} + \big( \frac{n\Theta_{n>0,m>0}}{\beta} \big) g_{n-1,m-1} \Big ) j_n(pq) \nonumber
\end{equation}
which can be computed once for all $p$ via Hankel transform. Furthermore, since the only dependence on $\hat{p}$ comes in via the $f_{n,m}, g_{n,m}$ we can precompute the scale dependence of these integrals as
\begin{equation*}
    \epsilon_{n,m}(p) = f_{n,m} \Gamma_{n,m} + n g_{n-1,m-1}\Gamma^{(-)}_{n,m}
\end{equation*}
where we have defined
\begin{align}
    \Gamma_{n,m} &= 4\pi  \int dq\ q^2 \ e^{\frac{1}{2}(\bk_1\cdot\bk_2)(X+Y)} \beta^n  q^{2m-n} j_n(pq) \\
    \Gamma^{(-)}_{n,m} &= 4\pi \Theta_{n>0,m>0} \int dq\ q^2 \ e^{\frac{1}{2}(\bk_1\cdot\bk_2)(X+Y)} \beta^{n-1}  q^{2m-n} j_n(pq).
\end{align}
For any given triangle $\bk_{1,2,3}$ these components can be pre-computed for all $p$ such that for a given $\bp$ evaluating $\mathcal{E}(\bk_1,\bk_2,\bp)$ reduces to simple matrix multiplication. 

\subsection{Recovering Tree Level Results}
\label{ssec:linear_theory}

The various kernels and infinite sums in the above being somewhat opaque, in this section we show how to recover the expected results at tree level by evaluating the first few summands at lowest order. We can begin by writing
\begin{align}
    \mathcal{E} = 4\pi \sum_{n} \frac{ (-1)^n}{n!} \int dq \ q^2 \Big( 1 &+ \frac{\bk_1 \cdot \bk_2}{2} (X+Y) \Big) \Big( \frac{k_1 k_2 Y}{2} \Big)^n \nonumber \\
    &(pq)^{-n} \Big( F_n - \frac{2 n \Theta_{n>0}}{k_1 k_2 Y} (pq)^2 G_{n-1}  \Big) j_n(pq) + \mathcal{O}(\PL^2).
\end{align}
Since each $X,Y$ carries one power of $\PL$, we can stop the sum at $n=2$ in order to recover linear theory:
\begin{align*}
    \mathcal{E}_{\rm lin} = 4\pi \int &dq\ q^2 \Bigg \{ \Big( \frac{\bk_1 \cdot \bk_2}{2} \Big) (X+Y)\ F_0 \ j_0(pq) \\
    &- (pq)^{-1} \Big(\frac{k_1 k_2 Y}{2} \Big) \ F_1 \ j_1(pq) + (pq)^{-2}\Big( \frac{k_1 k_2 Y}{2} \Big)^2 \Big( - \frac{4}{k_1 k_2 Y} G_{1} (pq)^2 \Big) j_2(pq) \Bigg \}.
\end{align*}
Using the definitions of $F_n$, $G_n$ and $X$ and $Y$ as well as the identity $j_1(x)/x = (j_0 + j_2)/3$ it is straightforward to recover
\begin{equation}
    \mathcal{E}_{\rm lin}^{\rm fin}(\bk_1, \bk_2, \bp) = -\frac{(\bk_1 \cdot \bp)(\bk_2 \cdot \bp)}{p^4} \PL(p).
\end{equation}
It is interesting to note that unlike in the case of the Zeldovich power spectrum, where $\bk_1 = -\bk_2$, recovering linear theory requires going to $n=2$. In general, in order to isolate contributions to $\mathcal{E}^{\rm fin}$ at a given order we note that we can write
\begin{equation*}
    \Gamma^{(-)}_{n,m} = \sum_r \frac{4\pi}{r!}  \int dq\ q^2 \ \left(\frac{1}{2}(\bk_1\cdot\bk_2)(X+Y)\right)^r \beta^{n(-1)}  q^{2m-n} j_n(pq)
\end{equation*}
such that the order-$N$ contributions to $\Gamma^{(-1)}_{n,m}$ come from $N = r + n (-1)$, and the corresponding contributions to $\mathcal{E}$ come from setting all other terms in the sum to zero.

We can now proceed to obtain the tree-level bispectrum in our numerical implementation. Pluggin in the linear contribution to $\mathcal{E}$ as derived above into Equation~\ref{eqn:ir_safe} we get to leading order
\begin{align}
    B^{\rm tree}(\bk_1,\bk_2) &= \mathcal{E}_{\rm lin}^{\rm fin}(\bk_1,\bk_1+\bk_2,-\bk_1-\bp) \mathcal{E}_{\rm lin}^{\rm fin}(\bk_2,\bk_1+\bk_2,-\bk_2-\bp) + ... \nonumber\\
    &= \Big(1 + \frac{\bk_1 \cdot \bk_2}{k_1^2} \Big) \Big(1 + \frac{\bk_1 \cdot \bk_2}{k_2^2} \Big) \PL(\bk_1) \PL(\bk_2) +  \textrm{cycl.} 
\end{align}
as expected.

\subsection{Alternative Angular Decomposition}
\label{ssec:alternative}

\begin{figure}
    \centering
    \includegraphics[width=\textwidth]{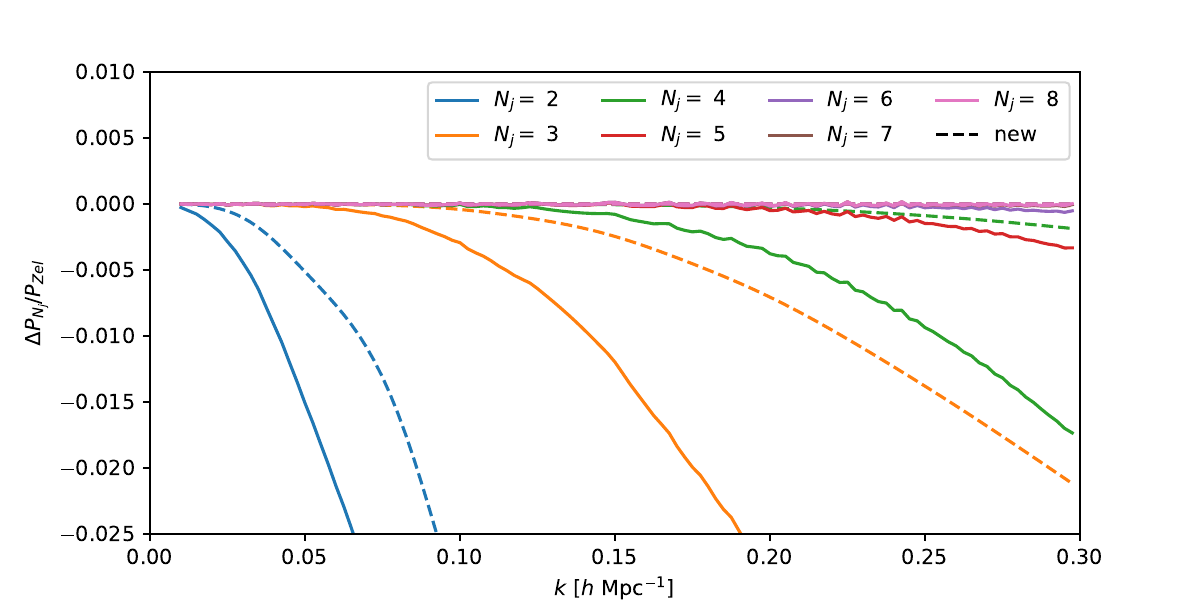}
    \caption{Convergence of the Hankel-transform series in Equation~\ref{eqn:new_hankels} compared to the fiducial method described in \S\ref{ssec:fiducial_method} in the case of the Zeldovich power spectrum. The series both converge to the correct values as the number $N_j$ of Hankel transforms increases, though Equation~\ref{eqn:new_hankels} does so substantially faster at almost all wavenumbers.}
    \label{fig:convergence_pk}
\end{figure}

It is also possible to perform the angular integral in $\mathcal{E}$ by directly Taylor-expanding the angular dependence in $Y(q)$. In this case we have
\begin{align}
    &\dtq e^{-i\bp\cdot\bq+\half k_{1,i} k_{2,j} A_{ij}(\bq)} \nonumber \\
    &= \dtq e^{-i\bp\cdot\bq+\half ((\bk_1\cdot\bk_2)X(q) + (\bk_1\cdot\hq)(\bk_2\cdot\hq)Y(q))} \nonumber \\
    &=\sum_{n=0}^\infty \frac{1}{n!} \dtq e^{-i\bp\cdot\bq+\half (\bk_1\cdot\bk_2)X(q)} k_{1,i_1} ... k_{1,i_n} k_{2,j_1} ... k_{2,j_n} \left(\frac{Y(q)}{2} \right)^n \hq_{i_1} ... \hq_{j_n} \nonumber \\
    &= \sum_{n=0}^\infty \frac{1}{n!} \dtq e^{-i\bp\cdot\bq+\half (\bk_1\cdot\bk_2)X(q)} k_{1,i_1} ... k_{1,i_n} k_{2,j_1} ... k_{2,j_n} \left(\frac{Y(q)}{2} \right)^n \sum_{m=0}^n  C_{n,m} [\mathcal{L}_{2m}(\hq)]_{i_1 ... j_n} \nonumber \\
    &= \sum_{n=0}^\infty \sum_{m=0}^n 4\pi \left( k_{1,i_1} ... k_{1,i_n} k_{2,j_1} ... k_{2,j_n} [\mathcal{L}_{2m}(\hp)]_{i_1 ... j_n} \right) \frac{(-1)^m}{n!}   \int dq\ q^2\ e^{\half(\bk_1 \cdot \bk_2)X(q)} C_{n,m} \left(\frac{Y(q)}{2} \right)^n j_{2m}(pq) \nonumber \\
    &= \sum_{m=0}^\infty \sum_{n=m}^\infty \frac{C_{n,m}}{n!} \left( k_{1,i_1} ... k_{1,i_n} k_{2,j_1} ... k_{2,j_n} [\mathcal{L}_{2m}(\hp)]_{i_1 ... j_n} \right) 4\pi (-1)^m \int dq\ q^2\ e^{\half(\bk_1 \cdot \bk_2)X(q)} \left(\frac{Y(q)}{2} \right)^n  j_{2m}(pq)   \nonumber  
\end{align}
where we have defined the coefficients
\begin{equation}
    \mu^{2n} = \sum_{m=0}^n C_{n,m} \mathcal{L}_{2m}(\mu).
\end{equation}
As in the previous method the final expression is a series of Hankel transforms whose integrands are independent of $\hp$ and proportional to powers of $Y(q)$, multiplied by the angular coefficient
\begin{equation}
    \mathcal{D}_{2m}(\hk_1, \hk_2, \hp) = \hk_{1,i_1} ... \hk_{1,i_n} \hk_{2,j_1} ... \hk_{2,j_n} [\mathcal{L}_{2m}(\hp)]_{i_1 ... j_n}.
\end{equation}
Here $\mathcal{L}_{2m}$ are the tensor equivalents of the Legendre polynomials (see e.g. ref.~\cite{Chen23}). The $\mathcal{D}_{2m}$ are polynomials in the dot products of the three unit vector arguments and can be worked out recursively using the recurrence relations of the Legendre polynomials. 

Let us consider the special case of the power spectrum $\bk = \bk_1 = -\bk_2 = \bp$. In this case the angular contraction is
\begin{equation*}
    \mathcal{D}(\hk,-\hk,\hk) = (-1)^n \hk_{i_1} ... \hk_{j_n} [\mathcal{L}_{2m}(\hk)]_{i_1 ... j_n} = (-1)^n
\end{equation*}
such that we can write
\begin{equation*}
    P_{\rm Zel}(k) = \sum_{m=0}^\infty 4\pi (-1)^m \int dq\ q^2\ e^{-\half k^2 X(q)}  j_{2m}(pq) F_m(q)
\end{equation*}
where
\begin{align}
    F_m(q) &= \sum_{n=m}^\infty \frac{C_{n,m}}{n!} \left(-\frac{k^2 Y(q)}{2} \right)^n \nonumber \\
    &= \frac{(4 m+1)  \Gamma \left(m+\frac{3}{2}\right)
   \, }{(2 m+1)\Gamma(2m+\frac32)}\ _1F_1\left(m+\frac{1}{2};2 m+\frac{3}{2};-\frac{k^2 Y(q)}{2}\right) \left(-\frac{k^2 Y(q)}{2} \right)^m.
   \label{eqn:new_hankels}
\end{align}
Note that $F_m$ is proportional to $Y^m$ to leading order, so this formula is a re-summation of the method presented above where terms with different powers were separated. Figure~\ref{fig:convergence_pk} shows the convergence of this sum as $m_{max}$ is increased---it is in excellent agreement with the fiducial method and indeed exhibits significantly faster convergence at large wavenumbers.

\section{IR Cancellation and Enhancement by BAO at 1-loop Order}
\label{app:ept_1loop}

Let us first consider the infrared properties of the kernels $Z_n$, i.e. their asymptotic limits when one or more momenta are small compared to the total momentum. In particular, if the momenta $\bp_{m}$ satisfy $p_m \ll k$ we have that
\begin{equation}
    Z_n(\bp_1, ..., \bp_m, \bk_1, ..., \bk_{n-m}) \rightarrow \frac{(n-m)!}{n!} \left( \frac{\bp_1 \cdot \bk}{p_1^2} \right) ... \left( \frac{\bp_m \cdot \bk}{p_m^2} \right) Z_{n-m}(\bk_1, ..., \bk_{n-m}),
\end{equation}
i.e. the kernels factor into ``shifts'' and lower-order kernels. Here $\bk = \bk_1 + ... + \bk_{n-m} \approx \bp_1 + ... + \bp_m + \bk_1 + ... + \bk_{n-m}$.

Let us now consider each contributing diagram to the 1-loop bispectrum. Since the IR limits of the kernels $Z_n$ reduce to shift terms involving long-wavelength modes $\bp_m$ we integrate over and lower order kernels, let us look specifically for diagrams whose IR limits approach piece of the tree level bispectrum proportional to $Z_2(\bk_1, \bk_2) \PL(k_1) \PL(k_2)$. All other IR contributions will be equivalent up to permutations of the external wavenumbers. Each of the diagrams we will evaluate in the IR limit is shown in the order below in Figure~\ref{fig:loop_diagrams}.

\subsection{The 114 contribution}

In this case we want to consider the diagram
\begin{align*}
    B_{114} &= \avg{ \delta^{(1)}(\bk_1) \delta^{(1)}(\bk_2) \delta^{(4)}(\bk_3) }' \\
    &= 12 \int_\bp Z_4(\bp, -\bp, \bk_1, \bk_2)\ \PL(p) \PL(k_1) \PL(k_2) \\
    &\supset \int_{\bp}^{k_{\rm IR}} \left( \frac{\bp \cdot \bk_{12}}{p^2} \right) \left( \frac{-\bp \cdot \bk_{12}}{p^2} \right) Z_2(\bk_1,\bk_2) \ \PL(p) \PL(k_1) \PL(k_2) \\
    &= \left( - k_{3,i} k_{3,j} \int_{\bp}^{k_{\rm IR}} \frac{p_i p_j}{p^4} \PL(p) \right) Z_2(\bk_1,\bk_2)\ \PL(k_1) \PL(k_2) \\
    &= - \half k_3^2 \Sigma^2_{\rm IR}\ 2 Z_2(\bk_1,\bk_2)\ \PL(k_1) \PL(k_2)
\end{align*}
where we have defined
\begin{equation}
    \Sigma^2_{\rm IR} = \third \int_{\bp}^{k_{\rm IR}} \frac{\PL(p)}{p^2}.
\end{equation}
and used that $\bk_{12} = - \bk_3$.

\subsection{The 123 contribution}

In this case we have the diagram
\begin{align*}
    B_{123} &= \avg{ \delta^{(1)}(\bk_1) \delta^{(2)}(\bk_2) \delta^{(3)}(\bk_3) }' \\
    &\supset 6 \int_{\bp} Z_2(\bp, \bk_2 - \bp) Z_3(\bp,\bk_2-\bp, \bk_1)\ \PL(p) \PL(|\bk_2 - \bp|) \PL(k_1) \\
    &\supset 12 \int_{\bp}^{k_{\rm IR}} Z_2(\bp, \bk_2 - \bp) Z_3(\bp,\bk_2-\bp, \bk_1)\ \PL(p) \PL(|\bk_2 - \bp|) \PL(k_1) \\
    &\rightarrow 2  \int_{\bp}^{k_{\rm IR}}  \left( \frac{\bp \cdot \bk_{2}}{p^2} \right) \left( \frac{\bp \cdot \bk_{12}}{p^2} \right) Z_2(\bk_1, \bk_2)\ \PL(p) \PL(|\bk_2 - \bp|) \PL(k_1) \\
    &=  -2 k_{2,i} k_{3,j} Z_2(\bk_1, \bk_2) \int_{\bp}^{k_{\rm IR}} \frac{p_i p_j}{p^4} \PL(p) \PL(|\bk_2 - \bp|) \PL(k_1) \\
    &\rightarrow - (\bk_2 \cdot \bk_3) \Sigma^2_{\rm IR} \  2 Z_2(\bk_1,\bk_2) \PL(k_1) \PL(k_2).
\end{align*}
In the third line, we have used that there are two IR poles in the integrand at $\bp = \bf{0}, \bk_2$, and that the integrand is invariant when shifting to the latter. There is an additional contribution to in $B_{213}$ with $\bk_1 \leftrightarrow \bk_2$, such that the total IR contribution is
\begin{equation}
    - (\bk_2 \cdot \bk_3 + \bk_1 \cdot \bk_3) \Sigma^2_{\rm IR} \  2 Z_2(\bk_1,\bk_2) \PL(k_1) \PL(k_2)= + k_3^2 \Sigma^2_{\rm IR} \  2 Z_2(\bk_1,\bk_2) \PL(k_1) \PL(k_2).
\end{equation}

\subsection{The 132 contribution}

Here we have the diagram
\begin{align*}
    B_{132} &= \avg{ \delta^{(1)}(\bk_1) \delta^{(3)}(\bk_2) \delta^{(2)}(\bk_3) }' \\
    &\supset 6 \int_{\bp} Z_2(\bk_1, \bk_2) Z_3(\bp,-\bp, \bk_2)\ \PL(p)  \PL(k_1) \PL(k_2) \\
    &\rightarrow   \int_{\bp}^{k_{\rm IR}}  \left( \frac{\bp \cdot \bk_{2}}{p^2} \right) \left( \frac{-\bp \cdot \bk_{2}}{p^2} \right)\ Z_2(\bk_1, \bk_2)\ \PL(p) \PL(k_1) \PL(k_2) \\
    &= \left(- \half k_2^2 \Sigma^2_{\rm IR} \right)\ 2 Z_2(\bk_1, \bk_2) \PL(k_1)\PL(k_2).
\end{align*}
Here again we can permute $\bk_{1,2}$ to get the total contribution
\begin{equation}
    \left(- \half (k_1^2 + k_2^2) \Sigma^2_{\rm IR} \right)\ 2 Z_2(\bk_1, \bk_2) \PL(k_1)\PL(k_2).
\end{equation}

\subsection{The 222 contribution}

Finally we have the diagram
\begin{align*}
    B_{222} &= \avg{ \delta^{(2)}(\bk_1) \delta^{(2)}(\bk_2) \delta^{(2)}(\bk_2) }' \\
    &\supset 8 \int_{\bp} Z_2(\bp, \bk_1 - \bp) Z_2(-\bp, \bk_2+\bp) Z_2(\bk_1-\bp,\bk_2+\bp)\ \PL(p)  \PL(|\bk_1-\bp|) \PL(|\bk_2+\bp|) \\
    &\rightarrow 2  \int_{\bp}^{k_{\rm IR}} \left( \frac{\bp \cdot \bk_{1}}{p^2} \right) \left( \frac{-\bp \cdot \bk_{2}}{p^2} \right) Z_2(\bk_1, \bk_2)\ \PL(p)  \PL(|\bk_1-\bp|) \PL(|\bk_2+\bp|) \\
    &\rightarrow  - (\bk_1 \cdot \bk_2) \Sigma^2_{\rm IR}\ 2 Z_2(\bk_1, \bk_2)\ \PL(k_1) \PL(k_2).
\end{align*}
Note that this diagram has three IR regions, identical to the full resummation discussed in \S\ref{ssec:ir_safe_integrand}--however, shifting to the other IR regions produces terms proportional to $P(k_3)$ instead, which are not degenerate with the terms discussed here but rather permutations thereof, so we can ignore them.

Combining the previous results we have that the full IR contribution the 1-loop bispectrum is given by
\begin{equation}
    \half \left(-k_3^2 + 2 k_3^2  - (k_1^2 + k_2^2) - 2 \bk_1 \cdot \bk_2  \right) \Sigma^2_{\rm IR}\ 2 Z_2(\bk_1, \bk_2) \PL(k_1) \PL(k_2) = 0,
\end{equation}
that is long-wavelength displacements do not contribute to the 1-loop bispectrum.

\subsection{Enhancement by the BAO}

In the above cancellations we have implicitly used that we can approximate
\begin{equation}
    \PL(|\bk + \bp|) = \PL(k) \left(1 + \mathcal{O}\left(\frac{p}{k}\right) \right).
\end{equation}
However, when the power spectrum contains an oscillatory component $P_w \sim \sin(r_s k)$ we have that the corrections are instead of order powers of $r_s$ times the signal, or enhanced by $r_s k$ compared to expectations. In other words, a mode $\bp$ which is in the infrared compared to $\bk$  can still have a large effect due to the size of $r_s$.

In the 1-loop bispectrum we can consider this enhancement by looking at the contributions due to $B_{123}$ and $B_{222}$ for which we have the IR contribution
\begin{align*}
    -2 \left( k_{2,i} k_{3,j} + k_{2,i} k_{1,j} \right) &Z_2(\bk_1, \bk_2) \int_{\bp}^{k_{\rm IR}} \frac{p_i p_j}{p^4} \PL(p) \PL(|\bk_2 - \bp|) \PL(k_1) \\
    &= 2 k_{2,i} k_{2,j} Z_2(\bk_1, \bk_2) \int_{\bp}^{k_{\rm IR}} \frac{p_i p_j}{p^4} \PL(p) \PL(|\bk_2 - \bp|) \PL(k_1)
\end{align*}
where we have dropped the $\bp$ in $\PL(|\bk_1 - \bp|)$ to focus on the effect on the BAO in $\PL(k_2)$. Isolating the BAO component in $\PL(\bk_2-\bp)$ then gives us, subtracting the $\bp \rightarrow 0$ limit by previous arguments
\begin{align*}
    k_{2,i} k_{2,j} &\ 2Z_2(\bk_1, \bk_2) \PL(k_1) \int_{\bp}^{k_{\rm IR}} \frac{p_i p_j}{p^4} \PL(p) ( P_w(|\bk_2 - \bp|) - P_w(k_2)) \\
    &= \half k_{2,i} k_{2,j} \ 2Z_2(\bk_1, \bk_2) \PL(k_1) \int_{\bp}^{k_{\rm IR}} \frac{p_i p_j}{p^4} \PL(p) ( P_w(|\bk_2 - \bp|) + P_w(|\bk_2 + \bp|) - 2 P_w(k_2)) \\
    &= \half k_{2,i} k_{2,j} \ 2Z_2(\bk_1, \bk_2) \PL(k_1) \int d^3\bx\ e^{-i\bk_2\cdot \bx} \xi_w(\bx)\ \int_{\bp}^{k_{\rm IR}} \frac{p_i p_j}{p^4} \PL(p)\ (2 \cos(\bp \cdot \bx) - 2).
\end{align*}
Since the BAO component of the correlation function is peaked at $x = r_s$ we see that any contribution with $p \ll 1/r_s$ is suppressed, but scales between $1/r_s$ and $k_2$ can contribute to the BAO despite being in the IR regime. Indeed, setting $\bx = r_s \hat{x}$ and performing the $\bp$ integral we can see that it yields precisely $A_{ij}(\bx) = X(r_s) \delta_{ij} + Y(r_s) \hat{x}_i \hat{x}_j$, such that the total contribution to the bispectrum is
\begin{equation}
    - \half k_2^2 \Sigma^2_{s} P_w(k_2)
\end{equation}
i.e. the leading order contribution to the damping of the BAO derived using the wiggle no-wiggle split in \S\ref{sec:ir_resummation}.

\section{Corrections to Saddle-Point in Power Spectrum}
\label{app:saddle_point}

Let us consider the integral
\begin{equation}
    P_{w,\rm Zel}(k) = \dtq e^{i\bk\cdot\bq - \half k_i k_j A_{ij}^{nw}(\bq)}\ \xi_w(q).
\end{equation}
To use the saddle-point approximation we make the Taylor expansion about $\bq = r_d \hq + (q - r_d) \hq$:
\begin{equation}
    A_{ij}^{nw}(\bq) = A_{ij}^{nw}(r_d \hq) + (q - r_d)\ \hq_a \partial_a A_{ij}^{nw}(r_d \hq) + \ldots\ .
\end{equation}
The first term gives us the leading order saddle-point result, i.e. exponential damping. The second term leads to a correction
\begin{equation}
    -\half k_i k_j \dtq e^{i\bk\cdot\bq - \half k_i k_j A_{ij}^{nw}(\bq)}\ (q - r_d)\ \hq_a \partial_a A_{ij}^{nw}(r_d \hq)\ \xi_w(q).
\end{equation}
Defining
\begin{equation}
    \xi^\ell_n(q) = \int \frac{dk\ k^{2+n}}{2\pi^2} P(k) j_\ell(kq)
\end{equation}
we have that the derivative above is given by
\begin{align}
\frac{\partial A^{nw}_{ij}(\bq)}{\partial q_k} = \frac25 \xi^{1,nw}_{-1}(q)  \hq_{(i} \delta_{jk)} -\frac45 \xi^{3,nw}_{-1}(q) \left( \frac52 \hq_i \hq_j \hq_k - \frac12 \hq_{(i} \delta_{jk)} \right)
\end{align}
and, working to leading order we get that the correction is
\begin{equation*}
    -\half k^2 \Big[ \frac23 \xi^1_{-1}(r_d) P^1_{w,0}(k) - \frac23 \left(\frac45 \xi^1_{-1}(r_d) - \frac65 \xi^3_{-1}(r_d)  \right) P^1_{w,2}(k)\Big].
\end{equation*}
In the above we have defined
\begin{equation}
    P^n_{w,\ell}(k) = 4\pi \int dq \ q^2 \ (q - r_d)^n\ \xi_w(q) \ j_\ell(kq).
\end{equation}
Roughly speaking these should be order the width of the (linear) BAO $r_D^n$. The cumulative correction to the damping is thus of order $r_D r_d \sigma^2_{r_d}$, where we have used that $\xi^1_{-1}(q) = q \sigma^2_q/ 3$ is related to the mean square density $\sigma^2_q$ in spheres of radius $q$---in practice we find that the effect is sufficiently small to be subdominant even to differences due to the choice of the wiggle no-wiggle split for the BAO, though its relevance is proportional to the width of the feature in the correlation function which can vary for e.g. primordial features.

\section{Angular Integral in Wiggle No-Wiggle Split}
\label{sec:derivation_xpy}

In this section we perform the angular integral in Equation~\ref{eqn:saddle_point_pk} to derive the damping form for the BAO component of the power spectrum with $\Sigma^2_s = X_s + Y_s$. In order to do so we will use the series and angular decomposition in \S\ref{ssec:alternative}. Let us take the special case where the Zeldovich integrand multiplies $\xi_w(q)$ such that $X,Y$ are isolated at their values at the peak $X_s, Y_s$. In the power spectrum for example this corresponds to the BAO component of the $b_1^2$ term
\begin{align}
    P_w^{NL}(k) &= \dtq e^{-i\bk\cdot\bq-\half k_{i} k_{j} A_{ij}(\bq)} \ \xi_w(q) \nonumber \\
    &\approx \sum_{n=0}^\infty \sum_{m=0}^n 4\pi \frac{(-1)^m}{n!}   \int dq\ q^2\ e^{-\half k^2 X_s} C_{n,m} \left(-\frac{k^2 Y_s}{2} \right)^n\ \xi_w(q) \ j_{2m}(kq) \nonumber \\
    &=  e^{-\half k^2 X_s} \sum_{n=0}^\infty \frac{1}{n!}\ \left(-\frac{k^2 Y_s}{2} \right)^n\ \sum_{m=0}^n C_{n,m}\ P_{w,2m}(k).
    \label{eqn:new_hankel_saddle}
\end{align}
Here we have defined
\begin{equation}
    P_{w,\ell} = 4\pi (-i)^\ell \int dq\ q^2 \ \xi_w(q)\ j_\ell(kq).
\end{equation}
Since $\xi_w$ is well localized at $r_s$, for any sufficiently large $k > r_s^{-1}$ wherever the integral has support we can make the asymptotic approximation that
\begin{equation}
    j_{2m}(kq) \approx \frac{1}{kq} \cos(kq - \pi/2 - m\pi) = \frac{(-1)^m}{kq} \cos(kq - \pi/2)\approx (-1)^m j_0(kq)
    \label{eqn:asymptotic}
\end{equation}
such that $P_{w,2m} \approx (-1)^m P_w$. Plugging this approximation back into Equation~\ref{eqn:new_hankel_saddle} and using $\sum_{m=0}^n C_{n,m} = 1$ yields
\begin{align}
    P_w^{NL}(k) = e^{-\half k^2 X_s} &\sum_{n=0}^\infty \frac{1}{n!} \left(-\frac{k^2 Y_s}{2} \right)^n \left( \sum_{m=0}^n C_{n,m} \right) P_{w}(k) \nonumber \\
    &= e^{-\half k^2 X_s} \sum_{n=0}^\infty \frac{1}{n!} \left(-\frac{k^2 Y_s}{2} \right)^n P_{w}(k) = e^{-\half k^2 \Sigma^2_s} P_w(k)
\end{align}
with $\Sigma^2_s = X_s + Y_s$ as desired.

We can also consider BAO contribtions with higher angular dependence in $\bq$ space. For example in the galaxy power spectrum we have linear terms proportional to
\begin{equation*}
    -\half k_i k_j A^w_{ij}(\bq) \sim k^2 \mathcal{L}_{0,2}(\mu) \int \frac{dp}{2\pi^2} \ j_{0,2}(pq)\ P_w(p),\quad i k_i U_i \sim k \mu \int \frac{dp\ p }{2\pi^2} \ j_{1}(pq)\ P_w(p)
\end{equation*}
which Fourier transform into $P_w$ at leading order. For simplicity let us consider the latter $i k_i U_i$ term: upon resumming the linear displacements this yields
\begin{align}
    P^w_{b_1} &= i k_a \dtq e^{-i\bk\cdot\bq - \half k_i k_j A_{ij}(\bq)} U_w(q) \hq_a \nonumber \\
    &= i k_a \sum_{n=0}^\infty \frac{\hk_{i_1} ...  \hk_{i_{2n}}}{n!} \dtq e^{-i\bk\cdot\bq - \half k^2 X(q)}\ U_w(q)  \left( \frac{-k^2 Y(q)}{2} \right)^n \hq_a \hq_{i_1} ... \hq_{i_{2n}} \nonumber \\
    &\approx i k_a e^{-\half k^2 X_s}  \sum_{n=0}^\infty \frac{\hk_{i_1} ...  \hk_{i_{2n}}}{n!}  \left( \frac{-k^2 Y_s}{2} \right)^n \dtq e^{-i\bk\cdot\bq}\ U_w(q)\  \hq_a \hq_{i_1} ... \hq_{i_{2n}} \nonumber
\end{align}
From here we can again decompose
\begin{equation}
    \mu^{2n + 1} = \sum_{m=0}^n C_{n,m}^{(1)} \mathcal{L}_{2m+1}(\mu)
\end{equation}
to write
\begin{equation*}
    \dtq e^{-i\bk\cdot\bq}\ U_w(q)\  \hq_a \hq_{i_1} ... \hq_{i_{2n}} = \sum_{m=0}^n 4\pi (-i)^{2m+1} C^{(1)}_{n,m} \int dq\ q^2\ U_w(q)\ j_{2m+1}(kq)\ \mathcal{L}_{2m+1}(\hk) %? }
\end{equation*}
and, using the asymptotic form for $j_{2m+1}$ as in Equation~\ref{eqn:asymptotic} and that $\mathcal{L}_\ell(1) = 1$ gives that $P^w_{b_1}$ is also damped by $\Sigma_s^2$, as are indeed all other BAO contributions by the same argument.

\section{Bispectrum of Biased Tracers}
\label{app:biased_tracers}

In this section we briefly sketch the calculation of the bispectrum of biased tracers in real space, leaving further details to future work. In order to simplify things we will work in the so-called shifted operator basis \cite{Schmittfull19} where the real-space galaxy overdensity is given by
\begin{equation}
    \delta_g(\bk) = \dtq e^{-i \bk \cdot (\bq + \Psi(\bq)}\ \left( b_1 \delta(\bq) + \frac{1}{2} b_2 (\delta^2(\bq) - \avg{\delta^2}) + b_s (s^2(\bq) - \avg{s^2}) + ... \right)\ .
\end{equation}
In order to compute the resummed galaxy bispectrum to leading order it is necessary to keep up to quadratic bias terms ($b_1, b_2, b_s$). Unlike in the Lagrangian bias basis there is no ``1'' term.

In order to compute the contributions from bias operators it is useful to first evaluate the second cumulant of the generating function generating function \cite{CLPT}
\begin{equation}
\mathcal{C} = \frac{1}{2}\langle \big( - i \bk_1 \cdot \Delta_{13} - i \bk_2 \cdot \Delta_{23} + \sum_n (\lambda_n \delta_n + a_{n,ij} s_{n,ij}) \big )^2 \rangle_c
\end{equation}
such that contributions due to each bias term can be obtained by taking derivatives of $\lambda_n$ and $a_n$. Here we have
\begin{align}
    \mathcal{C} =\ &\mathcal{C}_{\rm Zel} -i (\lambda_3 \bk_1 - \lambda_1 \bk_3) \cdot U_{13} -i (\lambda_3 \bk_2 - \lambda_2 \bk_3) \cdot U_{23} -i (\lambda_2 \bk_1 - \lambda_1 \bk_2) \cdot U_{12} \nonumber \\
    &-i (a_{3,ab} \bk_{1,i} - a_{1,ab} \bk_{3,i}) B_{13,iab} -i (a_{3,ab} \bk_{2,i} - a_{2,ab} \bk_{3,i}) B_{23,iab} -i (a_{2,ab} \bk_{1,i} - a_{1,ab} \bk_{2,i}) U_{12,iab} \nonumber \\
    &+ \lambda_1 \lambda_3 \xi_{13} + \lambda_2 \lambda_3 \xi_{23} + \lambda_1 \lambda_2 \xi_{12} + a_{1,ab} a_{3,cd} C_{13,abcd} + a_{2,ab} a_{3,cd} C_{23,abcd} + a_{1,ab} a_{2,cd} C_{12,abcd} \nonumber \\
    &+ (a_{1,ab} \lambda_3 + a_{3,ab} \lambda_1) E_{13,ab} + (a_{2,ab} \lambda_3 + a_{3,ab} \lambda_2) E_{23,ab} + (a_{1,ab} \lambda_2 + a_{2,ab} \lambda_1) E_{12,ab}\nonumber
\end{align}
where we have defined $U_{13} = \langle \delta_1 \Delta_{13} \rangle = - \langle \delta_1 \Psi_3 \rangle$ etc. The analagous due to the shear tensor ($C_{abcd}$, $E_{ab}$, $B_{iab}$) were defined in \cite{Kokron22}. From this it is clear the leading-order bispectrum terms decompose into separable functions of $\br$, $\bq$ and $\bq - \br$.

Proceeding to the bispectrum itself we see that the integrand is now the same as for Zeldovich matter but multiplied by
\begin{align}
&b_1^3 (-i (\bk_1 \cdot U_{13} + \bk_2 \cdot U_{23}) \xi_{12} - i (\bk_1 \cdot U_{12} - \bk_3 \cdot U_{23}) \xi_{13} +i (\bk_2 \cdot U_{12} + \bk_3 \cdot U_{13}) \xi_{23} \nonumber \\
&+ b_1^2 b_2 (\xi_{12} \xi_{13} + \xi_{13} \xi_{23} + \xi_{12} \xi_{23}) + 2 b_1^2 b_s (E_{12,ab} E_{13,ab} + E_{12,ab} E_{23,ab} + E_{13} E_{23})
\end{align}
The top line can be symmetrized by changing the order of $1,2,3$ and noting the sign flip upon inversion.

From the above we can see that the bispectrum can again be computed as a convolution over triplets of the functions
\begin{align*}
    \mathcal{E}_{U,\xi,E} = \dtq e^{-i\bp\cdot\bq+\half \bk_{1,i} \bk_{2,j} A_{ij}(\bq)}\ \big(U,\ \xi, \ E \big)(\bq) 
\end{align*}
These can be evaluated using the additional angular integrals
\begin{equation}
    I_n^m = \int d\Omega_{\hat q}\,
\left[ \hat k_{1,i} \hat k_{2,j} \lb \hat q_i \hat q_j - \delta^{\rm K}_{ij} \rb \right]^n (\hq_{i_1} \hq_{i_2} ... \hq_{i_m})\
e^{i \alpha \bp \cdot \hat q} .
\end{equation}
up to $m=2$. It is straightforward to see that these can be obtained by differentiating with respect to $\bp_i$.  For example we have that
\begin{equation}
    I_n^1 = \frac{1}{\alpha} \frac{d I_n(a,b,c,\alpha p)}{dp_i} = \frac{1}{\alpha} \Big( \frac{\partial I_n}{\partial b} \hk_{2,i} + \frac{\partial I_n}{\partial c} \hk_{1,i} + \alpha \frac{\partial I_n}{\partial p} \hp_i \Big)
\end{equation}
The derivatives with respect to $b,c$ are straightforward since $I_n$ are polynomials in these variables.

\bibliographystyle{JHEP}
\bibliography{main}
\end{document}